\documentclass[onecolumn,secnumarabic,amsmath,amssymb,balancelastpage,nofootinbib]{article}

\usepackage[e]{esvect}
\usepackage{color}         
\usepackage{graphics}      
\usepackage{graphicx}      
\usepackage{epsf}          
\usepackage{bm}            

\usepackage{natbib}
\usepackage{amssymb}
\usepackage{amsmath, esint}
\usepackage{mathrsfs}
\usepackage{framed}
\usepackage{bigints} 
\usepackage{enumitem}
\usepackage{pifont}
\usepackage[capbesideposition={left,center},facing=yes,capbesidewidth=8cm,capbesidesep=quad]{floatrow} 
\usepackage{setspace} 

\usepackage[none]{hyphenat} 

\usepackage[colorlinks=true]{hyperref}  

\definecolor{darkred}{rgb}{0.6,0,0}
\definecolor{darkgreen}{rgb}{0,0.5,0}
\definecolor{darkblue}{rgb}{0,0,0.6}
\hypersetup{ colorlinks,
linkcolor=darkblue,
filecolor=darkgreen,
urlcolor=darkgreen,
citecolor=darkred }

\newcommand{\del}[0]{\ensuremath{\vec{\nabla}}}
\newcommand{\sigmaT}[0]{\ensuremath{\overset{\text{\tiny$\leftrightarrow$}}{\sigma}}}
\newcommand{\tensor}[1]{\ensuremath{\overset{\text{\tiny$\leftrightarrow$}}{#1}}}

\begin{document}

\sloppy 



\title{\vspace*{-35 pt}\Huge{The Mass of the Gravitational Field}}
\author{Charles T. Sebens\\California Institute of Technology}
\date{January 10, 2019\\ arXiv v.2\\\vspace*{9 pt}Forthcoming in \emph{The British Journal for the Philosophy of Science}}

\maketitle
\vspace*{-15 pt}
\begin{abstract}

By mass-energy equivalence, the gravitational field has a relativistic mass density proportional to its energy density.  I seek to better understand this mass of the gravitational field by asking whether it plays three traditional roles of mass: the role in conservation of mass, the inertial role, and the role as source for gravitation.  The difficult case of general relativity is compared to the more straightforward cases of Newtonian gravity and electromagnetism by way of gravitoelectromagnetism, an intermediate theory of gravity that resembles electromagnetism.

\end{abstract}

\setstretch{1.1}
{\small \tableofcontents}
\setstretch{1.3}
\newpage

\section{Introduction}

By mass-energy equivalence ($\mathcal{E}=m c^2$), the gravitational field has a relativistic mass density proportional to its energy density.   I seek to better understand this mass of the gravitational field by asking whether it plays the traditional roles of mass---asking whether the gravitational field really acts like it has mass.

This paper is organized into sections focusing on four different physical theories: electromagnetism, Newtonian gravity, gravitoelectromagnetism, and general relativity.  For each theory, I ask whether the field has a mass playing any or all of the following three roles: the conservational role (in ensuring conservation of mass), the inertial role (in quantifying resistance to acceleration), and the gravitational role\footnote{This role of mass as source of gravitation is sometimes called the `active' gravitational role to distinguish it from the `passive' gravitational role of mass in quantifying the amount of force felt from a given gravitational field.  To limit the scope of this article, the passive gravitational role is not examined here.} (as source of gravitation).  Here I summarize the results.

The electromagnetic field possesses a mass playing all three of the above roles and thus serves as a useful point of comparison for analyzing the gravitational field.  The mass of the electromagnetic field plays the conservational role in ensuring that---although charged matter will generally gain and lose (relativistic) mass in interacting with the electromagnetic field---the total mass of field and matter is always conserved.  The fact that the mass of the electromagnetic field plays the inertial role is often illustrated in an indirect way: a charged body requires more force to accelerate than an uncharged body because there is additional mass in the electromagnetic field surrounding the charged body.  However, this effect is complicated by the fact that accelerated charged bodies emit electromagnetic radiation.  We can see the inertial role more directly by giving a force law that describes the reaction of the electromagnetic field to the forces exerted upon it by matter.  In general relativity, the mass of the electromagnetic field clearly plays the gravitational role.

In Newtonian gravity, the gravitational field does not possess a mass playing any of the above three roles.  One does not need to attribute any mass to the gravitational field to ensure conservation of mass.  However, one does need to attribute negative energy to the gravitational field in order to ensure conservation of energy.  There are a number of ways to do so, one of which \citet{maxwellgravity} found by analogy with electromagnetism.  The Newtonian gravitational field carries no momentum and thus has no mass playing the inertial role.  The Newtonian gravitational field does not act as a source for itself and thus has no mass playing the gravitational role.

Gravitoelectromagnetism is a theory of gravity that can be arrived at as an extension of Newtonian gravity (as was first done by \citealp{heaviside1893}) or as a limit of general relativity (as is done in some textbooks).  Gravitoelectromagnetism gets its name from the fact that the laws of the theory are structurally a very close match to the laws of electromagnetism.  Because of this close match, it is straightforward to show that in this theory the gravitational field possesses a mass playing the conservational and inertial roles.  The inertial role played by this mass can be seen directly in the force law for the field and indirectly in the fact that massive bodies are surrounded by clouds of negative field mass which make them easier to accelerate.  The force required to accelerate a body is also modified by the presence of gravitational radiation which carries away negative energy.  In gravitoelectromagnetism---as it is standardly presented---the gravitational field does not act as a source for itself and thus does not play the gravitational role.  However, one can modify the theory (making it nonlinear) so that the field is a source for itself.  The gravitational field around a planet is thus slightly weakened as it is now sourced by both the positive mass of the planet itself and the negative mass of its surrounding gravitational field.  Gravitoelectromagnetism is not nearly as well-known or well-studied as the other theories discussed in this paper, but I've included it as it serves as a useful bridge linking electromagnetism, Newtonian gravity, and general relativity.

In general relativity there are multiple mathematical objects that can be used to describe the flow of energy and momentum (related to the different energy densities available in Newtonian gravity), including the Weinberg and Landau-Lifshitz energy-momentum tensors.  I discuss these two tensors and call for another which better aligns with the three other physical theories discussed above.  However one describes the flow of energy, the mass of the gravitational field plays the conservational role.  It also plays the inertial role in a similar way to the mass of fields in the other theories.  The mass of the gravitational field appears to play the gravitational role, though it is difficult to say how it does so as this seems to depend on the way Einstein's field equations are written.  In my treatment of general relativity I adopt a field-theoretic approach to the theory (as is done in the textbooks of \citealp{weinbergGR} and \citealp{feynmanGR}) in order to stress the connections between general relativity and the three other theories.

The questions about the mass of the gravitational field pursued in this paper are relevant to a number of ongoing debates in the foundations of physics.  First, there has been much discussion about how to properly understand mass-energy equivalence, as expressed mathematically by Einstein's famous $\mathcal{E}=m c^2$.\footnote{See the references in appendix \ref{MassEnergyEquivSec}.}  Material bodies clearly possess mass and---by mass-energy equivalence---also possess an energy that is proportional to their mass.  Fields clearly possess energy and---by mass-energy equivalence---also possess a mass that is proportional to their energy.  But, do they really act like they possess such a mass?  That is the core question of this paper.  Second, there is an ongoing debate as to the ontological status of fields like the electromagnetic field and the gravitational field.  Are fields real and if so what kind of thing are they?  One argument that can be given for the reality of a particular field is that conservation of energy only holds if one takes the field to be a real thing that possesses energy.\footnote{See, e.g., (\citealp[ch.\ 5]{lange}; \citealp[p.\ 31]{frisch2005}; \citealp[sec.\ 4.2]{Lazarovici2017}).}  But, some are not convinced.  \citet{lange} contends that it is really the possession of proper mass by a field (not energy) which grounds the best argument for the electromagnetic field's reality.  I will not explicitly address the question of whether fields are real in the paper, but I think that everything I do to show that the mass possessed by fields acts just like the mass possessed by ordinary matter suggests that fields are just as real as matter.  Taking the gravitational field to be real in general relativity, one might wonder whether it is a field on spacetime or a part of spacetime.\footnote{See the references in footnotes \ref{findingthefield}, \ref{spin2toGR}, and \ref{manifoldapproach}.}  The former perspective will prove useful for our purposes here.  Third, philosophers of physics have recently put forward functionalist accounts of a number of important concepts, including probability, spacetime, and gravitational energy-momentum.\footnote{See (\citealp[ch.\ 4]{wallace2012}; \citealp{knoxF, baker2018}; \citealp{readforth}).}  According to a functionalist account of mass, a quantity would count as a mass provided it played certain roles like the three analyzed here.  Although I think it is of interest to determine whether these roles are played independent of any functionalist ambition, one could certainly build on the work done in this paper to develop a functionalist account of the mass of fields.  Fourth, there has been much debate as to how to properly understand the energy and momentum of the gravitational field in general relativity.\footnote{See the references in footnote \ref{GRreferences}.}  Asking about the mass of the gravitational field gives a slightly different angle on this well-studied problem (as the field's mass density is proportional to its energy density).  In section \ref{GRsection} I present the progress I've made.  However, the discussion here is restricted to the field-theoretic formulation of general relativity and the bearing of my conclusions on the theory's standard geometric formulation is left unsettled.

This paper is part of a larger research project on the mass of fields, including recent papers on the electromagnetic field and the Dirac field \citep{forcesonfields, howelectronsspin}.

\section{The Mass of the Electromagnetic Field}\label{sectionEM}

In the context of special relativity, each material body has a velocity-dependent relativistic mass (proportional to its energy).  That mass plays a number of different roles.  First, it plays the conservational role.  Globally, the sum of all relativistic mass never changes.  Locally, any change in a body's relativistic mass can be attributed to a (local) exchange of relativistic mass between that body and something else.  Second, relativistic mass plays the inertial role.  This mass quantifies the body's resistance to being accelerated (though the exact sense in which it does so is more complex than in pre-relativistic physics, as $\vec{F}$ is no longer equal to $m\vec{a}$).  Third, relativistic mass plays the gravitational role.  In general relativity, it is this mass which acts as a source for gravitation.  In relativistic contexts I will use `mass' as shorthand for `relativistic mass', as it is relativistic mass and not proper mass which most directly plays these three roles.\footnote{Although some other authors use `mass' as shorthand for relativistic mass (e.g., \citealp{bondispurgin}), many think that proper mass is more deserving of the title (see \citealp{okun1989}; \citealp[pp.\ 250--251]{taylorwheeler}; \citealp{lange2001}).}

According to mass-energy equivalence, if something has energy $\mathcal{E}$ it has mass $\mathcal{E}/c^2$.  The electromagnetic field thus possesses a mass proportional to its energy.  Upon first encountering the idea that the electromagnetic field has mass, one might think that this must be a very different sort of mass than the mass of an ordinary material body.  It is not.  The mass of the electromagnetic field plays all three of the above roles.  In this section I explain how it does so.  (See appendix \ref{MassEnergyEquivSec} for more on mass-energy equivalence and \citealp{forcesonfields} for more on the inertial role played by the electromagnetic field's mass.)

Before we start on all of that, let me pause to preempt a potential confusion.  When particle physicists discuss the `mass' of a given field, they are usually talking about a certain quantity which appears in the dynamical equations for the field and corresponds to the proper mass of the particle associated with that field.\footnote{It is this sense of field `mass' that is being used when authors describe the gravitational field of general relativity as a massless spin-two tensor field (see the references in footnote \ref{GRreferences}).}  In this sense, the electromagnetic field is massless because the photon has no proper mass (in contrast to, for example, the Dirac field).  That is not the sense of field `mass' which I am examining here.  When I talk about the mass of a field I am talking about the relativistic mass\footnote{One might wonder whether there is a proper mass of the field distinct from the proper mass of the particle associated with the field.  This is discussed in (\citealp[ch.\ 8]{lange}; \citealp[sec.\ 7]{forcesonfields}).} of the field, proportional to the field's energy.  Even though the photon has no proper mass, the electromagnetic field still has a relativistic mass density equal to its energy density divided by $c^2$.

\subsection{The Conservational Role}\label{secconsroleEM}

We must attribute energy to the electromagnetic field to ensure conservation of energy in electromagnetism---the energy of charged matter alone is not conserved.  As mass is proportional to energy, the mass of matter alone is similarly not conserved.  However, if we attribute mass to the electromagnetic field in proportion to its energy, the total mass of matter and field is conserved.

The laws of electromagnetism are Maxwell's equations \eqref{maxwellGEM} and the Lorentz force law \eqref{lorentzforcelawGEM}.  From these laws, one can derive an equation for the conservation of energy (Poynting's theorem),
\begin{equation}
\frac{\partial}{\partial t}\left[\frac{1}{8 \pi}\left(E^2+B^2\right)\right]+\vec{\nabla}\cdot \vec{S}_f=-\vec{f}_f\cdot\vec{v}^{\,q}_m
\ .
\label{energyconsEM}
\end{equation}
The first term gives the rate at which the energy of the electromagnetic field,
\begin{equation}
\rho^{\mathcal{E}}_f=\frac{1}{8 \pi}\left(E^2+B^2\right)\ ,
\label{energydensityEM}
\end{equation}
is changing.  (The $\mathcal{E}$ superscript on $\rho^{\mathcal{E}}_f$ indicates that this is the density of energy and the $f$ subscript indicates that it is a property of the electromagnetic field.)  The second term in \eqref{energyconsEM} describes the rate at which field energy flows out of a volume in terms of the Poynting vector (the energy flux density),
\begin{equation}
\vec{S}_f= \frac{c}{4\pi} \vec{E} \times \vec{B}\ .
\label{PoyntingEM}
\end{equation}
The righthand side of \eqref{energyconsEM} gives the rate at which energy is transferred from matter to field (per unit volume).  This rate is expressed in terms of the work done by the Lorentz force density $\vec{f}_f$ as $-\vec{f}_f\cdot\vec{v}^{\,q}_m=-\vec{J}_m\cdot\vec{E}$, where $\vec{J}_m=\rho_m^q\vec{v}^{\,q}_m$ is the current density, $\rho_m^q$ is the charge density, and $\vec{v}^{\,q}_m$ is the velocity field describing the flow of charge.  (The $q$ superscript indicates that these quantities describe the flow of charge as opposed to the flow of mass or energy and the $m$ subscript indicates that these are properties of matter.)  Note that attributing energy to the electromagnetic field, as in \eqref{energydensityEM}, obviates the need to attribute potential energy to matter which, for example, increases as oppositely charged bodies are pulled away from one another.\footnote{\citet[ch.\ 5]{lange} discusses why the non-instantaneous nature of electromagnetic interactions makes the inclusion potential energy insufficient to achieve conservation of energy in electromagnetism (i.e., why one must attribute an energy density to the electromagnetic field as in \eqref{energydensityEM}).  On p.\ 119 he mentions that such an introduction of field energy (replacing potential energy) is not necessitated in Newtonian gravitation, though in section \ref{sectionNG} I will adopt such a picture as it better aligns with the other theories considered here.}

By mass-energy equivalence ($\mathcal{E}=m c^2$), the electromagnetic field has a mass density equal to its energy density \eqref{energydensityEM} divided by $c^2$,
\begin{equation}
\rho_f=\frac{1}{8 \pi c^2}\left(E^2+B^2\right)\ .
\label{massdensityfield}
\end{equation}
We can divide the above energy conservation equation \eqref{energyconsEM} everywhere by $c^2$ to arrive at an equation for the conservation of mass,
\begin{equation}
\frac{\partial \rho_f}{\partial t}+\vec{\nabla}\cdot \vec{G}_f=\frac{-\vec{f}_f\cdot\vec{v}^{\,q}_m}{c^2}
\ ,
\label{massconsEM}
\end{equation}
where $\vec{G}_f$ is the momentum density of the electromagnetic field, equal to the energy flux density divided by $c^2$:
\begin{equation}
\vec{G}_f= \frac{\vec{S}_f}{c^2} = \frac{1}{4\pi c} \vec{E} \times \vec{B}\ .
\label{momentumEM}
\end{equation}
Thinking of momentum as relativistic mass times velocity, we can write $\vec{G}_f$ as $\rho_f \vec{v}_f$ by introducing a velocity $\vec{v}_f$ to describe the flow of the electromagnetic field's mass,\footnote{See (\citealp{poincare1900}; \citealp[p.\ 373]{kraus1953}; \citealp{geppert1965, arora1967}; \citealp[sec.\ 14.2.1]{bornwolf}; \citealp[p.\ 122]{MTW}; \citealp{forcesonfields}).}
\begin{equation}
\vec{v}_f=\frac{\vec{G}_f}{\rho_f}=\frac{\vec{S}_f}{\rho^{\mathcal{E}}_f}=2c\frac{\vec{E} \times \vec{B}}{\left(E^2+B^2\right)}
\ .
\label{fieldvelocity}
\end{equation}
This velocity cannot exceed the speed of light (as it attains its maximum value, $c$, when $\vec{E}$ and $\vec{B}$ are perpendicular and equal in magnitude).

\subsection{The Inertial Role}\label{secinertialroleEM}

The inertial role of mass is its role in quantifying the amount of acceleration a body experiences in response to a given force.  The fact that the electromagnetic field possesses such a mass can be seen in an indirect but compelling way by observing that it is harder to accelerate a charged body (which carries along some field mass with it) than an otherwise similar uncharged body (which carries none).  The apparent inertial mass of the charged body is larger because both the mass of the field and the mass of matter are resisting acceleration.  But, this is only one of two effects modifying the way a charged body reacts to forces (as compared to an uncharged body).  Charged bodies also experience radiation reaction forces when they are accelerated because they emit electromagnetic radiation which carries away energy and momentum.\footnote{For an introduction to radiation reaction, see (\citealp{pearle1982}; \citealp[ch.\ 11]{griffiths}; \citealp[ch.\ 16]{jackson}).\label{emradreaction}}  The radiation reaction force may point opposite the acceleration (making the body even harder to accelerate) or it may point in some other direction.  The total force on the body resulting from these two distinct effects can be called the `field reaction'.\footnote{This terminology follows \citep[sec.\  11.2]{griffiths}.  Griffiths divides the electromagnetic field surrounding an accelerated charge into a `velocity field' which stays with the charge as it moves and an `acceleration field' which radiates off to infinity.  He goes on to use this division of the field into two parts to explain the difference between field reaction and radiation reaction: `As the particle accelerates and decelerates energy is exchanged between it and the velocity fields, at the same time as energy is irretrievably radiated away by the acceleration fields. ... if we want to know the recoil force exerted by the fields on the charge, we need to consider the \emph{total} power lost at any instant, not just the portion that eventually escapes in the form of radiation.  (The term ``radiation reaction'' is a misnomer.  We should really call it \emph{field reaction}. ...)'}  For an ordinary macroscopic charged object, the field reaction is insignificant because the amount of energy in the electromagnetic field is negligible.  However, for subatomic particles (like the electron) the field reaction becomes quite important.

\begin{figure}[htb]\centering
\includegraphics[width=7 cm]{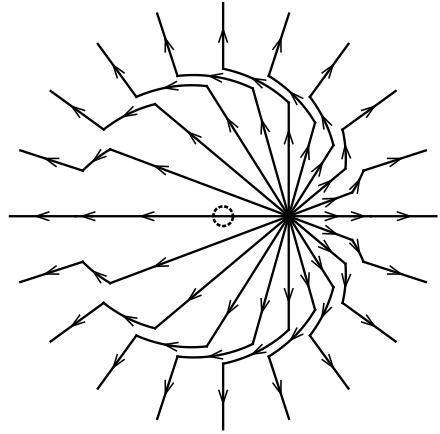}
\caption{This figure shows the electric field lines around a positive charge that was initially at rest and then for a brief period quickly accelerated to half the speed of light.  The dashed circle indicates the particle's initial position.  Figures like this are discussed in \citep[sec.\ 5.7]{purcell}.}
\label{acceleratedcharge}
\end{figure}

As a simple example of field reaction, consider a spherical positively charged body which is acted upon by an applied force that uniformly accelerates it from rest over a very short period of time and then ceases.  The electric field around such a body---after the acceleration has finished but before the radiation has escaped too far---is depicted in figure \ref{acceleratedcharge}.  Before and after the acceleration, the charge is surrounded by an outwardly directed electric field (and after by a magnetic field as well).  Interpolating between these pre- and post-acceleration electric fields is a quite differently oriented electric field resulting from the period of acceleration.  During the period of acceleration, a field like this passes through the body and---as can be seen by looking at the direction of the field lines in the figure---the net electric force points opposite the direction of acceleration, making it more difficult to accelerate the charge.  Both of the effects mentioned in the previous paragraph contribute to making the charge more difficult to accelerate:  The applied force must accelerate not just the mass of body itself, but the mass of the field surrounding it as well.  Also, the applied force must provide the energy which is radiated away in electromagnetic waves.

The above approach to understanding the inertial role played by the electromagnetic field is indirect.  The mass of the field should quantify resistance to acceleration of the thing which possesses that mass---the field itself.  Using the field velocity given above \eqref{fieldvelocity}, it can be shown that the field's mass does play this role.  Consider the conservation of momentum equation for electromagnetism,
\begin{equation}
-\vec{f}_f=\frac{\partial}{\partial t}\left(\rho_f \vec{v}_f\right)-\vec{\nabla}\cdot \sigmaT_{\!f}\ ,
\label{forcelawEM}
\end{equation}
where $-\vec{f}_f$ is the force density exerted by matter upon the field (equal and opposite the force exerted by the field upon matter) and $\sigmaT_{\!f}$ is the momentum flux density tensor for the electromagnetic field (also known as the Maxwell stress tensor),
\begin{equation}
\sigmaT_{\!f}=\frac{1}{4\pi}\vec{E}\otimes\vec{E}+\frac{1}{4\pi}\vec{B}\otimes\vec{B}-\frac{1}{8\pi}\left(E^2+B^2\right)\tensor{I} \ .
\label{maxwellstresstensor}
\end{equation}
Upon integrating \eqref{forcelawEM} over a volume, the left side gives the force exerted on the field in that volume, the first term on the right gives the rate at which the momentum of the field in that volume is changing, and the second term on the right gives the rate at which field momentum is leaving that volume.

It is not standard to speak, as I just did, of forces acting upon the electromagnetic field.  But, I think it is helpful to do so.  The response of the electromagnetic field to what I have described as a force exerted by matter has the same form as the relativistic Eulerian force law giving the response of matter (modeled as a continuum) to the equal and opposite force exerted by the field upon matter,
\begin{equation}
\vec{f}_f=\frac{\partial}{\partial t}\left(\rho_m \vec{v}_m\right)-\vec{\nabla}\cdot \sigmaT_{\!m}\ ,
\label{forcelawmatterEM}
\end{equation}
where $\rho_m$ is the relativistic mass density, $\vec{v}_m$ is the velocity of mass flow, and $\sigmaT_{\!m}$ is the momentum flux density tensor for matter.  Comparing \eqref{forcelawEM} and \eqref{forcelawmatterEM}, we see that \eqref{forcelawEM} is an Eulerian force law for the electromagnetic field.  In that equation the mass of the electromagnetic field quantifies resistance to acceleration in just the same way that the mass of matter quantifies resistance to acceleration in \eqref{forcelawmatterEM}.  To better understand the inertial role of the field's mass, we could also analyze the Lagrangian forms of these force laws (which use the material derivative $\frac{D}{Dt}$), though we will not do so here (see \citealp{forcesonfields}).

\subsection{The Gravitational Role}

In general relativity, the mass of the electromagnetic field acts as a source of gravitation in just the same way that the mass of matter does.  In fact, the electromagnetic field is considered to be `matter' in the broad way the term is often used in the context of general relativity.

\section{The Mass of the Gravitational Field in Newtonian Gravity}\label{sectionNG}

In Newtonian gravity the gravitational field does not possess a mass playing any of the three roles we just went through.  This should come as no surprise.  Mass-energy equivalence was never a part of Newton's theory of gravity.  Still, it is worthwhile to see exactly how the gravitational field fails to play these roles for future comparison between Newtonian gravity and more advanced theories of gravity.

Let us consider how the Newtonian gravitational field interacts with a continuous distribution of matter.  The following equation describes how mass acts as a source for the gravitational field:
\begin{equation}
 \del \cdot \vec{g}  = - \nabla^2 \phi = - 4 \pi G \rho_m
 \ .
\label{sourceNG}
\end{equation}
Here $\phi$ is the gravitational potential and $\vec{g}$ is the gravitational field, related to $\phi$ by $\vec{g}=-\del\phi$.  The density of force exerted upon matter by the gravitational field is
\begin{equation}
\vec{f}_g=\rho_m \vec{g}=-\rho_m \del\phi
\label{forceonmatterNG}
\end{equation}
The $g$ subscript indicates that a quantity pertains to the gravitational field (for example, $\vec{f}_g$ is the force exerted by the gravitational field).  As we will be considering three distinct theories of gravitation, be aware that expressions for such quantities will change.

The mass density $\rho_m$ that appears in \eqref{sourceNG} and \eqref{forceonmatterNG} is, of course, the density of ordinary non-relativistic mass.  In the context of Newtonian gravity, we will use the term `mass' to refer to this kind of mass.  In the next section, we will shift back to using `mass' for relativistic mass.

\subsection{The Conservational Role}\label{secconsroleNG}

In Newtonian gravity, there is no need to attribute mass to the gravitational field in order to ensure conservation of mass.  The (non-relativistic) mass of matter is itself conserved
\begin{equation}
\frac{\partial \rho_m}{\partial t} =-\del\cdot(\rho_m \vec{v}_m)
\ ,
\label{continuitymatterNG}
\end{equation}
in contrast to the (relativistic) mass of matter in electromagnetism---which is only conserved in conjunction with the (relativistic) mass of the electromagnetic field, see \eqref{massconsEM}.  The gravitational field does not have a mass playing the conservational role.

However, the gravitational field can be attributed an energy density to ensure conservation of energy.  From \eqref{sourceNG}, \eqref{forceonmatterNG}, and \eqref{continuitymatterNG}, we can derive a conservation of energy equation for gravity similar to \eqref{energyconsEM},
\begin{equation}
\frac{\partial}{\partial t}\left(\frac{-g^2}{8 \pi G}\right)+\vec{\nabla}\cdot \vec{S}_g=-\vec{f}_g\cdot\vec{v}_m
\ ,
\label{energyconsNG}
\end{equation}
where
\begin{equation}
\rho^{\mathcal{E}}_g=\frac{-g^2}{8 \pi G}=\frac{-|\del\phi|^2}{8 \pi G}\ 
\label{energydensityNG}
\end{equation}
is interpreted as the energy density of the gravitational field and $\vec{S}_g$ as the Poynting vector for the gravitational field (giving the gravitational energy flux density),\footnote{This Poynting vector appears in (\citealp[eq.\ 5.11, assuming $\rho_m=0$]{synge1972}; \citealp[eq.\ 10]{noonan1984}).}
\begin{equation}
\vec{S}_g=\frac{1}{4\pi G}\left(\phi \frac{\partial}{\partial t} \del \phi + \vec{v}_m\phi\nabla^2 \phi \right)
\ .
\label{PoyntingNG}
\end{equation}
In words, \eqref{energyconsNG} says that the rate at which the energy of the gravitational field in a volume changes plus the rate at which gravitational energy leaves that volume is equal to the rate at which energy is transferred from matter to the gravitational field within that volume.

The energy density of the Newtonian gravitational field in \eqref{energydensityNG} is quite similar in form to the energy density of the electromagnetic field \eqref{energydensityEM}.  However, because of the difference in sign the gravitational energy density is always negative.  Noting the similarities between gravity and electromagnetism, \citet[part IV]{maxwellgravity} tentatively proposed the above expression for the energy density of the gravitational field.  Maxwell was troubled by the fact that this energy density is negative.  He thought that because `energy is essentially positive', it would be `impossible for any part of space to have negative intrinsic energy' and thus that space must possess an `enormous [positive] intrinsic energy' which \eqref{energydensityNG} describes deficits of.  We need not share this particular concern.  Still, the idea that gravitational energy may be negative raises questions.  This feature of gravitational energy will be of special interest to us in the next two theories of gravity where, by mass-energy equivalence, we will be dealing with negative mass.

The Poynting vector for Newtonian gravity \eqref{PoyntingNG} looks odd in comparison to the Poynting vector for electromagnetism \eqref{PoyntingEM}.  The first oddity to note is that $\vec{S}_g$ is not expressed purely in terms of field variables, the velocity of matter $\vec{v}_m$ appears---as might $\rho_m$ if \eqref{PoyntingNG} were rewritten using \eqref{sourceNG}.  The second oddity is that $\vec{S}_g$ is written in terms of the gravitational potential $\phi$ and cannot be rewritten solely in terms of the gravitational field $\vec{g}$.  This should raise red flags that the Poynting vector may be gauge-dependent.  Indeed it is.  Should you shift $\phi$ everywhere by a constant $s$, $\vec{g}$ will be unchanged but $\vec{S}_g$ will acquire an additional contribution of $\frac{s}{4\pi G}\left( \frac{\partial}{\partial t} \del \phi + \vec{v}_m\nabla^2 \phi \right)$.  However, this additional contribution is divergenceless---as can easily be seen using \eqref{sourceNG} and \eqref{continuitymatterNG}.  Thus, the additional contribution does not change the flux of energy out of any closed surface and does not appear in \eqref{energyconsNG}.\footnote{The possibility of adding divergenceless terms to the electromagnetic Poynting vector is discussed in (\citealp{lange2001}; \citealp[ch.\ 5]{lange}; \citealp[p.\ 347, footnote 1]{griffiths}; \citealp[sec.\ 6.7 and 12.10]{jackson}).}

It is worth noting that although we will continue to use the expressions for energy density and energy flux in \eqref{energydensityNG} and \eqref{PoyntingNG}, they are not unique.  Consider the alternative energy density of 
\begin{equation}
\rho^{\mathcal{E}}_{g}=\frac{1}{2}\rho_m \phi=\frac{\phi\nabla^2\phi}{8 \pi G}\ ,
\label{energydensityNG2}
\end{equation}
which, unlike \eqref{energydensityNG}, is a gauge-dependent function of the gravitational potential not expressible in terms of the gravitational field alone.  This density can be interpreted in a number of ways: as an energy density of the field, an interaction energy density of matter and field, or a potential energy density of matter.  I will treat it as an energy density of the field, like \eqref{energydensityEM} and \eqref{energydensityNG}.  One can rewrite the conservation of energy equation \eqref{energyconsNG} using this new energy density and a suitably altered Poynting vector,\footnote{This Poynting vector appears in \citep{bondi1962}---restricted to empty space where $\rho_m=0$.}
\begin{equation}
\frac{\partial}{\partial t}\left( \frac{1}{2}\rho_m \phi \right)+\vec{\nabla}\cdot \left[\frac{1}{4\pi G}\left(\frac{1}{2}\phi  \frac{\partial}{\partial t}  \del \phi -\frac{1}{2} \frac{\partial \phi}{\partial t}\del \phi  + \vec{v}_m\phi\nabla^2 \phi\right)\right]=-\vec{f}_g\cdot\vec{v}_m
\ .
\label{energyconsNG2}
\end{equation}
As will become relevant in section \ref{GRsection}, it is also possible to form energy densities by combining contributions from \eqref{energydensityNG} and \eqref{energydensityNG2}.\footnote{Such alternative expressions for gravitational energy density are discussed in (\citealp{peters1981}; \citealp[sec.\ 1.3]{ohanianGR}; \citealp[box 13.4]{thorneblandford}).}  The general form of such an energy density is
\begin{equation}
\rho^{\mathcal{E}}_{g}=\alpha\frac{ -g^2}{8 \pi G}+ (1-\alpha)\frac{1}{2}\rho_m \phi=\alpha\frac{- |\del\phi|^2}{8 \pi G}+(1-\alpha)\frac{\phi\nabla^2\phi}{8 \pi G}\ ,
\label{energydensityNG3}
\end{equation}
where $\alpha$ can be varied.  The conservation of energy equation for such an energy density is
\begin{align}
&\frac{\partial}{\partial t}\left(\alpha\frac{- g^2}{8 \pi G}+ (1-\alpha)\frac{1}{2}\rho_m \phi \right)
\nonumber
\\
&\quad+\vec{\nabla}\cdot \left[\frac{1}{4\pi G}\left(\frac{1+\alpha}{2}\:\phi  \frac{\partial}{\partial t}  \del \phi + \frac{\alpha-1}{2}\:\frac{\partial \phi}{\partial t}\del \phi  +\vec{v}_m\phi\nabla^2 \phi\right)\right]=-\vec{f}_g\cdot\vec{v}_m
\ .
\label{energyconsNG3}
\end{align}

\subsection{The Inertial Role}

Just as there is no need to attribute mass to the gravitational field in order to ensure conservation of mass, there is also no need to attribute momentum to the gravitational field in order to ensure conservation of momentum.   The integral of the force density \eqref{forceonmatterNG} over all of space is zero and thus the momentum of matter is conserved globally.  Local conservation of momentum can be expressed by an equation much like \eqref{forcelawEM},
\begin{equation}
-\vec{f}_g = -\vec{\nabla}\cdot \sigmaT_{\!g}\ ,
\label{forcelawNG}
\end{equation}
where $\sigmaT_{\!g}$ is the momentum flux density tensor for the gravitational field,\footnote{This tensor appears in (\citealp[eq.\ 11]{chandrasekhar1969}; \citealp[eq.\ 3.2]{synge1972}; \citealp[eq.\ 39.18]{MTW}; \citealp[eq.\ 4]{noonan1984}; \citealp[eq. 2.4]{giulini1997}; \citealp[eq.\ 5.61]{straumann2004}; \citealp[box 13.4]{thorneblandford}).} 
\begin{equation}
\sigmaT_{\!g}=\frac{-1}{4\pi G}\vec{g}\otimes\vec{g}+\frac{1}{8\pi G}g^2\,\tensor{I}
\ .
\label{momfluxNG}
\end{equation}
Interpreting \eqref{forcelawNG} analogously to \eqref{forcelawEM}, the equation states that any change in the momentum of matter in a volume is balanced by a flow of momentum into or out of that volume (as captured by the righthand side).  Because this balance is exact, there is never any accumulation of momentum in the gravitational field and thus the field has no momentum density (though it does have a momentum flux density since momentum flows through it).  This is in contrast to the electromagnetic field where $\vec{f}_f$ does not exactly balance $\vec{\nabla}\cdot \sigmaT_{\!f}$ and momentum can be transferred to or from the field (not merely through it).  The fact that momentum does not accumulate in the gravitational field can be seen as a result of the instantaneous nature of gravitational interactions between bodies.  Because the gravitational field has no momentum, it does not have any mass playing the inertial role.

Newtonian gravity is usually understood as a theory in which material bodies directly exert forces upon one another.  But, the above comparison with electromagnetism suggests a different picture.  Material bodies do not exert forces directly upon one another but instead these forces are mediated (though not delayed) by the gravitational field.  The gravitational field exerts forces on matter---given by \eqref{forceonmatterNG}---and matter exerts equal and opposite forces upon the gravitational field---the lefthand side of \eqref{forcelawNG}.  On the righthand side of \eqref{forcelawNG}, $\sigmaT_{\!g}$ acts as a stress tensor describing the forces exerted within the field (the field-on-field forces).\footnote{In \citep{forcesonfields} I distinguished between the momentum flux density for the electromagnetic field and the true stress tensor (which is not the Maxwell stress tensor).  We need not make such a distinction in the context of Newtonian gravity since the $\rho_g (\vec{v}_g \otimes \vec{v}_g)$ term that would differentiate the two tensors is zero (because the gravitational field has no mass).}  The fact that the left and right sides of \eqref{forcelawNG} are equal means that the net force on the field in any region of space is zero.  Thus, the field acts somewhat like an idealized massless string, which transmits forces but does not acquire momentum because the forces that act upon any bit of string are never unbalanced.

An example may help to illustrate the above understanding of forces in Newtonian gravity.  Suppose there are two massive bodies separated from each other by some distance.  Each will experience a force from the gravitational field directed towards the other body and each will exert a force on the gravitational field directed away from the other body.  These forces on the gravitational field will be balanced by tensile forces in the gravitational field connecting the forces exerted by the two bodies.\footnote{Such `gravitational tensions' are mentioned by \citet[sec.\ 4]{kaplan2009} in the context of gravitoelectromagnetism.  The idea of tensions in the electromagnetic field is common, dating back to Faraday's work (see, e.g., \citealp[pp.\ 187, 271--273]{whittaker1}; \citealp[sec.\ 5.6]{MTW}; \citealp[sec.\ 5]{forcesonfields}).}  This is similar to what happens in the electrostatic case of two opposite charges held in place some distance from one another.  Each experiences an electric force from the electromagnetic field directed towards the other body and each exerts a force on the electromagnetic field directed away from the other body.  Tensile forces in the electromagnetic field connect the forces exerted by the two bodies upon the field.  In this scenario, the electromagnetic field carries no momentum---as can be seen immediately from \eqref{momentumEM}, noting that there is no magnetic field.

When discussing the inertial role of field mass in electromagnetism we began by addressing the problem indirectly, asking whether it is more difficult to accelerate a charged body from rest than an uncharged body.  Similarly, one could ask if it is any more difficult to accelerate a body along with the gravitational field surrounding it than it would be to accelerate the body without its gravitational field (imagining gravity to be turned off).  Of course, it is no more difficult.  The gravitational field around a body does not contribute to its apparent inertial mass in Newton's theory of gravity.

\subsection{The Gravitational Role}

In \eqref{sourceNG}, the only source for gravity is the mass of matter, $\rho_m$.  One could imagine modifying this equation so that gravity acts as a source for itself, adding to the righthand side the energy density of the gravitational field divided by $c^2$.\footnote{A number of authors have considered such a variant of Newtonian gravitation, including \citet{einstein1912}; \citet{geroch1978}; \citet{peters1981}; \citet{visser1989}; \citet{giulini1997}; \citeauthor{jefimenko2000} (\citeyear{jefimenko2000}, \citeyear{jefimenko2006}); \citet{franklin2015}.}  In light of mass-energy equivalence, this seems like a natural move.  But, mass-energy equivalence is not part of Newtonian gravity and so we will delay consideration of such a move until the next section where we examine a theory of gravity that includes mass-energy equivalence.  For now, looking at Newtonian gravity as it is, the gravitational field has no mass playing the gravitational role.

\section{The Mass of the Gravitational Field in Gravitoelectromagnetism}\label{sectionGEM}

Gravitoelectromagnetism is an intermediate theory of gravity that serves as a useful stepping stone in moving from Newtonian gravity to general relativity and is thus sometimes included in textbooks on general relativity.  Gravitoelectromagnetism is not nearly as empirically successful as general relativity, but it contains some significant improvements over Newtonian gravitation: the theory accurately predicts Lense-Thirring precession and introduces a finite speed at which gravity propagates (because the theory includes gravitational waves which act like electromagnetic waves).  Studying gravitoelectromagnetism allows us to bring together what we've learned about mass in the electromagnetic and gravitational fields on our way to understanding the mass of the gravitational field in general relativity.  Gravitoelectromagnetism can be arrived at either as an extension of Newtonian gravity modeled off electromagnetism or as a weak-field slow-velocity approximation to general relativity.\footnote{For more on gravitoelectromagnetism as an extension of Newtonian gravity, see (\citealp{heaviside1893}; \citealp[ex.\ 7.2]{MTW}; \citeauthor{jefimenko2000}, \citeyear{jefimenko2000}, \citeyear{jefimenko2006}).  For more on gravitoelectromagnetism as a limit of general relativity and on the empirical observation of gravitomagnetic effects, see (\citealp{forward1961, braginsky1977}; \citealp[sec.\ 4.4a]{waldGR}; \citealp{harris1991, DSX1991, jantzen1992, ciufolini1995, maartens1998, mashhoon1999,mashhoon2001}; \citeauthor{mashhoon2001b}, \citeyear{mashhoon2001b}, \citeyear{mashhoon2007}; \citealp{clark2000, straumann2000, tartaglia2003}; \citealp[app.\ 17A]{hobsonGR}; \citealp{kaplan2009, keppel2009}; \citealp[sec.\ 3.4 and 4.7]{ohanianGR}; \citealp{bakopoulos2017}).  In this literature there is wide variation regarding notation and definitions.  I see this chaos as a license to define things as I see fit.  I've chosen to define the gravitoelectric and gravitomagnetic fields $\vec{E}_g$ and $\vec{B}_g$ so that they have the same units as $\vec{E}$ and $\vec{B}$.  This choice makes the expressions for energy density and momentum flux density particularly simple.  As compared to (\citeauthor{jefimenko2000}, \citeyear{jefimenko2000}, \citeyear{jefimenko2006}), my gravitoelectric field is $\frac{1}{\sqrt{G}}$ times his gravitational field $\vec{g}$ and my gravitomagnetic field is $\frac{c}{\sqrt{G}}$ times his cogravitational field $\vec{K}$.  As compared to \citep[app.\ 17A]{hobsonGR}, my gravitoelectric field is $\frac{1}{\sqrt{G}}$ times theirs and my gravitomagnetic field is $\frac{c}{4\sqrt{G}}$ times theirs.  As compared to (\citealp{mashhoon1999,mashhoon2001}; \citeauthor{mashhoon2001b}, \citeyear{mashhoon2001b}, \citeyear{mashhoon2007}), my gravitoelectric field is $\frac{-1}{\sqrt{G}}$ times theirs and my gravitomagnetic field is $\frac{-1}{2\sqrt{G}}$ times theirs.}  I present the first method here and the second in appendix \ref{GEMfromGR}.

If the relevant velocities are sufficiently small, the laws of electromagnetism, \eqref{maxwellGEM} and \eqref{lorentzforcelawGEM}, reduce to the laws of electrostatics,
\begin{align}
\mbox{\sc Elect}&\mbox{\sc rostatics}
\nonumber
\\
\vec{\nabla}\cdot\vec{E}&=4\pi \rho^q_m
\nonumber
\\
\vec{\nabla}\times\vec{E}&=0
\label{maxwellGS}
\\
\vec{f}_f &= \rho^q_m \vec{E}\ ,
\label{lorentzforcelawES}
\end{align}
because there is no appreciable magnetic field produced by moving charges.\footnote{From these laws of electrostatics, one could derive a conservation of energy equation similar to \eqref{energyconsNG} (with $\phi$ replaced by the electric potential $V$).  In so doing, a Poynting vector for the electric field would be introduced with the same defects as \eqref{PoyntingNG}.  As was the case for the Newtonian gravitational field, the energy density of the electric field would not be unique.  For example, the energy density $\frac{1}{2}\rho^q_m V$, like \eqref{energydensityNG2}, could be used in place of $\frac{E^2}{8 \pi}$ from \eqref{energydensityEM}  (\citealp[eq.\ 2.43]{griffiths}; \citealp[eq.\ 1.53]{jackson}; \citealp{peters1981}).}  These laws closely resemble the laws of Newtonian gravity, \eqref{sourceNG} and \eqref{forceonmatterNG} along with $\del\times\vec{g}=0$ (which follows from the fact that $\vec{g}$ is the gradient of a scalar field),
  \begin{align}
  \mbox{\sc Grav}&\mbox{\sc itostatics}
\nonumber
\\
\vec{\nabla}\cdot\vec{E}_g&=-4\pi \sqrt{G} \rho_m
\nonumber
\\
\vec{\nabla}\times\vec{E}_g&=0
\label{maxwellGS}
\\
\vec{f}_g &=\sqrt{G} \rho_m \vec{E}_g \ .
\label{lorentzforcelawGS}
\end{align}
Here we've replaced $\vec{g}$ by the rescaled $\vec{E}_g=\frac{1}{\sqrt{G}}\vec{g}$ (the `gravitoelectric field') and titled the theory `gravitostatics' to suggest that it is the low-velocity limit of a deeper theory.  The difference between gravitostatics and electrostatics is just that the charge density which generates the field has been replaced by $-\sqrt{G}$ times the mass density and the charge density used for calculating the force exerted by the field has been replaced by $\sqrt{G}$ times the mass density (these differing signs ensure that gravity is attractive and not repulsive).\footnote{We can thus think of mass as a kind of `gravitational charge'.  Or, following Coulomb, we could go the other way and think of electric charge as a kind of `electrical mass' \citep[p.\ 79]{roller1954}.}  We can extend this replacement procedure to full electromagnetism in order to get a theory of gravitoelectromagnetism in which moving mass produces a `gravitomagnetic field' just as moving charge produces a magnetic field.  (Of particular importance for applications of the theory is the fact that rotating planets produce gravitomagnetic fields which are measurable, but weak enough that their effects don't spoil the successes of Newtonian gravity.)  Here are the Maxwell equations and Lorentz force laws for the two theories,\footnote{Here I treat both fields as interacting only with matter and ignore any interactions between the electromagnetic and gravitational fields.} placed side-by-side to facilitate quick comparison:
\begin{align}
\mbox{\sc Electr}&\mbox{\sc omagnetism} & \mbox{\sc Gravito}&\mbox{\sc electromagnetism}
\nonumber
\\
\vec{\nabla}\cdot\vec{E}&=4\pi \rho^q_m & \vec{\nabla}\cdot\vec{E}_g&=-4\pi \sqrt{G} \rho_m
\nonumber
\\
\vec{\nabla}\times\vec{E}&=-\frac{1}{c}\frac{\partial \vec{B}}{\partial t} & \vec{\nabla}\times\vec{E}_g&=-\frac{1}{c}\frac{\partial\vec{B}_g}{\partial t}
\nonumber
\\
\vec{\nabla}\cdot\vec{B}&=0 & \vec{\nabla}\cdot\vec{B}_g&=0
\nonumber
\\
\vec{\nabla}\times\vec{B}&=\frac{4\pi}{c}\vec{J}_m+\frac{1}{c}\frac{\partial \vec{E}}{\partial t} & \vec{\nabla}\times\vec{B}_g&=-\frac{4\pi}{c}\sqrt{G}\vec{G}_m+\frac{1}{c}\frac{\partial \vec{E}_g}{\partial t}
\label{maxwellGEM}
\\
\vec{f}_f &= \rho^q_m \left(\vec{E} + \frac{1}{c}\vec{v}^{\,q}_m \times\vec{B} \right) & \vec{f}_g &=\sqrt{G} \rho_m \left(\vec{E}_g + \frac{4}{c}\vec{v}_m \times\vec{B}_g \right)\ .
\label{lorentzforcelawGEM}
\end{align}
Moving further away from Newtonian gravity, I will take the $\rho_m$ which appears in the laws of gravitoelectromagnetism to be the relativistic mass density of matter and henceforth resume the use of `mass' as shorthand for `relativistic mass'.  The momentum density of matter is $\vec{G}_m = \rho_m\vec{v}_m$.  Just as the electric and magnetic fields are referred to together as the electromagnetic field, the gravitoelectric and gravitomagnetic fields will be referred to together as the gravitational field (which is much less of a mouthful than `the gravitoelectromagnetic field').

Note that \eqref{maxwellGEM} and \eqref{lorentzforcelawGEM} include one important deviation from the recipe of substitution given above: there is a factor of 4 which accompanies the gravitomagnetic force on matter in $\vec{f}_g$ but not the magnetic force on matter in $\vec{f}_f$.  One would not have guessed this factor of 4 from the comparison with electromagnetism (and for that reason it does not appear in all presentations of gravitoelectromagnetism\footnote{See, e.g., (\citeauthor{jefimenko2000}, \citeyear{jefimenko2000}, \citeyear{jefimenko2006}).}).  The factor of 4 arises when you derive gravitoelectromagnetism as an approximation to general relativity (appendix \ref{GEMfromGR}) and is important to the theory's applications \citep[sec.\ 2.3]{straumann2000}.  As we will see, this disanalogy with electromagnetism leads to a violation of momentum conservation.  So, at times it will be helpful to compare gravitoelectromagnetism (as presented above) with an alternative theory in which the troublesome factor of 4 is not present.

\subsection{The Conservational Role}\label{secconsroleGEM}

From the gravitoelectromagnetic Maxwell equations and Lorentz force law, one can derive an equation for the conservation of energy analogous to Poynting's theorem in electromagnetism \eqref{energyconsEM},
\begin{equation}
\frac{\partial}{\partial t}\left[\frac{-1}{8 \pi}\left(E_g^2+B_g^2\right)\right]+\vec{\nabla}\cdot \vec{S}_g=-\vec{f}_g\cdot\vec{v}_m
\ .
\label{energyconsGEM}
\end{equation}
In this equation the aforementioned factor of 4 is irrelevant since the gravitomagnetic force drops out when $\vec{f}_g$ is dotted with $\vec{v}_m$.  The energy density of the gravitational field which appears in \eqref{energyconsGEM} is
\begin{equation}
\rho^{\mathcal{E}}_g=-\frac{1}{8 \pi}\left(E_g^2+B_g^2\right)\ ,
\label{energydensityGEM}
\end{equation}
and the Poynting vector for the gravitational field is\footnote{This Poynting vector appears in (\citealp{heaviside1893}; \citealp[eq.\ 6-2.41]{jefimenko2000}; \citealp[eq.\ 3.17]{mashhoon2007}).}
\begin{equation}
\vec{S}_g= - \frac{c}{4\pi} \vec{E}_g \times \vec{B}_g\ .
\label{PoyntingGEM}
\end{equation}

Dividing \eqref{energyconsGEM} by $c^2$ yields an equation for the conservation of mass similar to \eqref{massconsEM},
\begin{equation}
\frac{\partial \rho_g}{\partial t}+\vec{\nabla}\cdot \vec{G}_g=\frac{-\vec{f}_g\cdot\vec{v}_m}{c^2}
\ ,
\label{massconsGEM}
\end{equation}
where the mass density of the field is
\begin{equation}
\rho_g=-\frac{1}{8 \pi c^2}\left(E_g^2+B_g^2\right)\ ,
\label{massdensityGEM}
\end{equation}
and $\vec{G}_g$ is the momentum density arrived at by diving the Poynting vector by $c^2$,
\begin{equation}
\vec{G}_g=- \frac{1}{4 \pi c} \vec{E}_g \times \vec{B}_g\ .
\label{momdensGEM1}
\end{equation}
We can introduce a velocity for the gravitational field,
\begin{equation}
\vec{v}_g=\frac{\vec{G}_g}{\rho_g}=\frac{\vec{S}_g}{\rho^{\mathcal{E}}_g}=2c\frac{\vec{E}_g \times \vec{B}_g}{\left(E_g^2+B_g^2\right)}\ ,
\label{fieldvelocityGEM}
\end{equation}
which takes exactly the same form as \eqref{fieldvelocity} and is similarly capped at $c$.  Note that the mass and energy of the gravitational field are negative.  This is an important disanalogy with electromagnetism, the consequences of which will be explored shortly.

Before proceeding to consider the inertial role of the gravitational field's mass in gravitoelectromagnetism, let's briefly discuss the connection between the above equations for gravitoelectromagnetism and the corresponding equations for Newtonian gravity.  The energy density in \eqref{energydensityGEM} is a straightforward extension of the energy density in \eqref{energydensityNG} to incorporate the new gravitomagnetic field.  The Poynting vector in \eqref{PoyntingGEM} looks quite different from \eqref{PoyntingNG} and does not have the same defects.  Still, the divergence of $\vec{S}_g$---which appears in \eqref{energyconsGEM} and determines the flux of gravitational energy through any closed surface---will agree with the Newtonian expression in the appropriate limit.\footnote{Starting with $\vec{S}_g$ from \eqref{PoyntingGEM}, the divergence is
\begin{equation}
\del\cdot\vec{S}_g= - \frac{c}{4\pi} \left(\vec{B}_g\cdot(\del\times\vec{E}_g) - \vec{E}_g \cdot (\del \times \vec{B}_g)\right)
\ .
\end{equation}
In the Newtonian limit the curl of $\vec{E}_g$ is zero \eqref{maxwellGS}.  We can use the gravitoelectromagnetic Maxwell equations \eqref{maxwellGEM} to expand the curl of $\vec{B}_g$,
\begin{equation}
\del\cdot\vec{S}_g= \frac{1}{4\pi} \vec{E}_g \frac{\partial \vec{E}_g}{\partial t}   - \sqrt{G} \rho \vec{E}_g \cdot \vec{v}_m
\ .
\end{equation}
Using \eqref{sourceNG} and \eqref{continuitymatterNG}, this becomes
\begin{equation}
\del\cdot\vec{S}_g=\del\cdot\left[\frac{1}{4\pi G}\phi  \frac{\partial}{\partial t}  \del \phi + \frac{1}{4\pi G}\vec{v}\phi\del^2 \phi\right]
\ ,
\end{equation}
in agreement with \eqref{PoyntingNG}.}

\subsection{The Inertial Role}\label{secinertialroleGEM}

As before, one can approach the question of whether the gravitational field has a mass playing the inertial role indirectly by asking how difficult it is to accelerate a massive body along with its gravitational field as compared to accelerating the same body with gravity turned off.  In gravitoelectromagnetism, the fact that massive bodies are accompanied by clouds of negative field mass makes them easier to accelerate.  However, as in electromagnetism, there is a second effect to consider: radiation reaction.  In gravitoelectromagnetism, accelerating massive bodies gain energy from the emission of negative energy gravitational radiation.

Consider the simple case of applying an external force to accelerate a massive spherical body from rest.  As in section \ref{secinertialroleEM}, we can see that there will be a field reaction force opposing that acceleration by looking at the field lines.  Because Maxwell's equations are essentially unchanged, figure \ref{acceleratedcharge} will accurately depict the gravitoelectric field around a body that was temporarily accelerated from rest provided we flip the arrows on the field lines.  The field reaction force will thus point along the direction of acceleration, making the body easier to accelerate than it otherwise would be.

It is a problematic feature of the theory that the energy of gravitational waves is negative.  A physical system could potentially gain arbitrarily large amounts of energy by sending out waves of negative energy into empty space.  This peculiar feature of gravitoelectromagnetism is removed in general relativity where---although the energy of the gravitational field can still be negative---the energy carried away by gravitational waves is positive.

Let us now consider the inertial role directly, asking how hard it is to accelerate the gravitational field itself.  From \eqref{maxwellGEM} and \eqref{lorentzforcelawGEM} you can calculate an Eulerian force law for the gravitational field,
\begin{equation}
-\left[ \sqrt{G} \rho_m \left(\vec{E}_g + \frac{1}{c}\vec{v}_m \times\vec{B}_g \right) \right] = \frac{\partial}{\partial t}(\rho_g \vec{v}_g) - \vec{\nabla}\cdot \sigmaT_{\!g}\ ,
\label{forcelawGEM}
\end{equation}
where $\rho_g \vec{v}_g$ is the momentum density of the gravitational field and $\sigmaT_{\!g}$ is the momentum flux density tensor
\begin{equation}
\sigmaT_{\!g}=\frac{-1}{4\pi}\vec{E}_g\otimes\vec{E}_g+\frac{-1}{4\pi}\vec{B}_g\otimes\vec{B}_g+\frac{1}{8\pi}\left(E_g^2+B_g^2\right)\tensor{I}
\ ,
\label{momfluxGEM1}
\end{equation}
which clearly reduces to the Newtonian expression \eqref{momfluxNG} when the gravitomagnetic field is negligible.  The key difference between \eqref{forcelawGEM} and the conservation of momentum equation for electromagnetism \eqref{forcelawEM} is that the force exerted upon the gravitational field by matter appearing on the left side of \eqref{forcelawGEM} is not equal and opposite the force exerted by the gravitational field upon matter \eqref{lorentzforcelawGEM}.  Because of this violation of Newton's third law, momentum is not conserved in gravitoelectromagnetism (as the theory has been formulated above).  However, if the factor of 4 were not included in \eqref{lorentzforcelawGEM} (a possibility discussed earlier) then the forces would balance and momentum would be conserved in the theory.

Considering the theory without the factor of 4, \citet[ch.\ 8]{jefimenko2006} argues that Newton's third law does not hold in gravitoelectromagnetism (though conservation of momentum does).  An apparent violation of Newton's third law: when there are just two massive bodies in motion, the forces each one feels may not be equal and opposite.  If forces are understood to be exerted by each body on the other, this would amount to a violation of Newton's third law.  But, if we instead think of forces as exerted by the gravitational field upon matter and by matter back upon the gravitational field, then from \eqref{lorentzforcelawGEM} (without the 4) and \eqref{forcelawGEM} it is clear that Newton's third law is upheld.  The same maneuver can be used to save Newton's third law from a similar challenge in electromagnetism \citep{forcesonfields}.

\subsection{The Gravitational Role}\label{secgravroleGEM}

There is an inconsistency in gravitoelectromagnetism.  As the gravitational field exerts forces upon matter, energy will be exchanged between field and matter at a rate of $\vec{f}_g\cdot\vec{v}_m$ \eqref{energyconsGEM}.  By mass-energy equivalence, this means that the total mass of matter will be changing over time.  However, by evaluating $\del\cdot(\del\times\vec{B}_g)=0$ using the gravitoelectromagnetic Maxwell equations \eqref{maxwellGEM} you can derive an equation requiring conservation of mass for matter alone---just as one can derive the conservation of charge from the electromagnetic Maxwell equations.  One way to rid the theory of this inconsistency is to say that the mass density which appears in the theory's laws, \eqref{maxwellGEM} and \eqref{lorentzforcelawGEM}, is the density of proper mass, not relativistic mass.  Since proper mass will not change as matter gains and loses energy in gravitoelectromagnetic interactions, the proper mass of matter alone would be conserved.  However, this response takes gravitoelectromagnetism farther from general relativity, where it is relativistic mass (not proper mass) that acts as the source of gravitation.  Another way to cure this inconsistency is to modify the gravitoelectromagnetic Maxwell equations so that the mass of the gravitational field itself acts as a source for gravity (thereby having the mass of the gravitational field play all three of its rightful roles, not just two):\footnote{\citeauthor{jefimenko2000} (\citeyear{jefimenko2000}, \citeyear{jefimenko2006}) considers making such a modification, though he focuses primarily on gravitostatics (Newtonian gravity).}
\begin{align}
\mbox{\sc Gravito}&\mbox{\sc electromagnetism (with self-source terms)}
\nonumber
\\
\vec{\nabla}\cdot\vec{E}_g&=-4\pi \sqrt{G} (\rho_m+ \rho_g)
\nonumber
\\
\vec{\nabla}\times\vec{E}_g&=-\frac{1}{c}\frac{\partial\vec{B}_g}{\partial t}
\nonumber
\\
 \vec{\nabla}\cdot\vec{B}_g&=0
\nonumber
\\
\vec{\nabla}\times\vec{B}_g&=-\frac{4\pi}{c}\sqrt{G}\left(\vec{G}_m+\vec{G}_g\right)+\frac{1}{c}\frac{\partial \vec{E}_g}{\partial t}
\label{maxwellGEM2}
\\
\vec{f}_g &=\sqrt{G} \rho_m \left(\vec{E}_g + \frac{4}{c}\vec{v}_m \times\vec{B}_g \right)\ .
\label{lorentzforcelawGEM2}
\end{align}
The aforementioned inconsistency is removed as the equation for conservation of energy \eqref{energyconsGEM} is unchanged and $\del\cdot(\del\times\vec{B}_g)=0$ now yields a conservation law for the total mass of matter and field,
\begin{equation}
\frac{\partial \rho_m}{\partial t}+\frac{\partial \rho_g}{\partial t} =-\del\cdot(\rho_m \vec{v}_m)-\del\cdot(\rho_g \vec{v}_g)
\ .
\label{continuityGEM}
\end{equation}
However, these altered gravitoelectromagnetic equations give rise to an additional Lorentz-force-like term, $\sqrt{G} \rho_g \left(\vec{E}_g + \frac{1}{c}\vec{v}_g \times\vec{B}_g \right)$, in the conservation of momentum equation \eqref{forcelawGEM} so that now the theory violates conservation of momentum with or without the troublesome factor of 4 in \eqref{lorentzforcelawGEM2}.  One could imagine making further alterations to the laws of gravitoelectromagnetism in order to alleviate all such problems, but I will not explore that avenue here.\footnote{See the references in footnote \ref{spin2toGR} for more on this sort of problem as it arises in the context of general relativity (in particular, \citealp[p.\ 186]{MTW}).}

In gravitoelectromagnetism the total mass of a planet, including the gravitational field surrounding it, is less than the mass of the matter that composes the planet alone.  This is somewhat strange.  But, it is a strangeness that we should get used to: it will remain in general relativity\footnote{This effect is mentioned often in discussions of general relativity, e.g., (\citealp{arnowitt1960}; \citealp[sec.\ 3.1, 7.6, and 8.2]{weinbergGR}; \citealp[p.\ 467]{MTW}; \citealp{ohanian2010}).} and was already present (for energy but not mass) in Newtonian gravity.  Because it is this total mass $\rho_m+ \rho_g$, and not $\rho_m$ alone, which acts as the source of gravitation in \eqref{maxwellGEM2}, the gravitational field far away from a planet will be a bit weaker than you'd expect from \eqref{sourceNG} or \eqref{maxwellGEM}.  This effect is generally small.  For example, the mass of the earth's gravitational field is about $-4.2\times10^{-10}$ times the mass of the earth itself.\footnote{\citet[p.\ 70]{weinbergGR} gives an estimate of this contribution which appears to be off by a factor of two.}

\section{The Mass of the Gravitational Field in General Relativity}\label{GRsection}

The theory of general relativity is normally presented as a theory of spacetime geometry in which gravity is represented by the curvature of spacetime and not by a field on spacetime.  Such a geometric approach to general relativity makes it difficult to pose the questions we've been asking of the previous theories as it appears, \emph{prima facie}, that there is no gravitational field to ponder the mass of.  One might argue that even though there is no separate field defined on spacetime, there exists some aspect of the spacetime which deserves the name `gravitational field', such as the connection $\Gamma^{\alpha}_{\ \beta\gamma}$, the Riemann tensor $R^{\alpha}_{\ \beta\gamma\delta}$, or the metric $g_{\mu\nu}$.\footnote{The merits of these options are assessed in \citep[sec.\ 4]{lehmkuhl2008} and references therein.\label{findingthefield}}  However, a field which is part of spacetime looks quite different from the fields on spacetime that we have been evaluating in the preceding sections.

In order to more easily build on what we've learned so far, it will be helpful to focus on an alternative formulation of general relativity in which gravity is treated as a field on spacetime.  I will call this the field-theoretic approach to contrast it with the geometric approach described above.\footnote{Finding appropriate terminology here is difficult as there are a range of subtly different views about the status of spacetime geometry and gravitational field in general relativity.  \citet{lehmkuhl2008} carves things up differently.  According to Lehmkuhl, a `field interpretation' of general relativity `claims that the geometry of spacetime can be reduced to the behavior of gravitational fields'.  This category is meant to include the field-theoretic approach alongside others.  On the field-theoretic approach, the flat background spacetime is of course independent of the gravitational field and not reduced to it.  But, the metric $g_{\mu\nu}$---which is interpreted as specifying the geometry of spacetime on the geometric approach---is straightforwardly determined by the gravitational field according to \eqref{decomposition}.}  The field-theoretic approach to general relativity has been developed in a number of sources including the textbooks of \citet{weinbergGR} and \citet{feynmanGR}.\footnote{For more on the field-theoretic approach to general relativity, see (\citealp{kraichnan1955,gupta1957,deser1970}; \citealp[boxes 7.1, 17.2, and 18.1]{MTW}; \citealp{straumann2000}; \citealp[sec.\ 5]{lehmkuhl2008}; \citealp[ch.\ 3]{ohanianGR}).  Some of these authors (as well as \citealp[p.\ 171]{weinbergGR}; \citealp[pp.\ x--xv and ch.\ 6]{feynmanGR}) present a field-theoretic way of `deriving' general relativity: start with a spin-two tensor field theory sourced by the energy-momentum tensor of matter (not including gravity) and then in clearing up the inconsistency problems that will arise (similar to those presented in section \ref{secgravroleGEM}) introduce a sequence of appropriate additional source terms (whereby gravity acts as a source for gravity) until the theory is rendered consistent.  (For a recent critical discussion of this sort of derivation, see \citealp{padmanabhan}; cf. \citealp{pittsschieve2007}.)\label{spin2toGR}}  On this way of understanding the theory, what used to be interpreted as the spacetime metric, $g_{\mu\nu}$, is broken up into a background spacetime metric (which we will take to be flat) and a (spin-two) tensor field on that spacetime,
\begin{equation}
g_{\mu\nu}=\eta_{\mu\nu}+h_{\mu\nu}\ ,
\label{decomposition}
\end{equation}
where $\eta_{\mu\nu}$ is the Minkowski metric and $h_{\mu\nu}$ is a potential specifying the state of the field (like $\phi$ in section \ref{sectionNG} or $A_g^\mu$ in appendix \ref{GEMfromGR}).\footnote{One could alternatively take $g_{\mu\nu}$ (not $h_{\mu\nu}$) to give the state of the gravitational field, thinking of the background spacetime as a metric-less manifold.  This idea is discussed in \citep{earman1987, maudlin1988, hoefer1996, rey2013}.  Such an approach is distinct from the field-theoretic and geometric approaches described above.\label{manifoldapproach}}  The laws that govern $g_{\mu\nu}$ (the Einstein field equations) are unchanged.

The decomposition in \eqref{decomposition} is central to the linear approximation to general relativity where $h_{\mu\nu}$ is assumed to be small.  Here no such assumption is made.  The proponents of the field-theoretic approach see it as a way of formulating general relativity which is (at least largely) equivalent to the theory's more standard geometric formulation.\footnote{See (\citealp[p.\ 165]{weinbergGR}; \citealp[ch.\ 6 and 8]{feynmanGR}).}  The field-theoretic approach is presented as preferable to the geometric approach since it makes gravity look more like the other physical fields that appear in particle physics and thus serves as a useful framework for developing a quantum theory of gravity.

On the field-theoretic approach, the background spacetime $\eta_{\mu\nu}$ is unobservable---other decompositions of the metric into background spacetime and field would work just as well since it is the total metric $g_{\mu\nu}$ which matters for deriving predictions from the theory.  \citet[p.\ 286]{waldGR} has criticized the introduction of such a decomposition of the metric into unobservable flat background spacetime and dynamical field as against the `spirit' of general relativity, writing that: `Such additional structure would be completely counter to the spirit of general relativity, which views the spacetime metric as fully describing all aspects of spacetime structure and the gravitational field.'  This charge of heresy could certainly be challenged---debate over the spirit of general relativity should be permitted among those studying the foundations of physics.  However, this is not the only problem facing the field-theoretic approach.  The approach has also been criticized for functioning poorly when deviations from the Minkowski metric are large---in particular when one is attempting to describe the physics of black holes---and for being only applicable to spacetimes which have a simply connected Euclidean topology.\footnote{See (\citealp[sec.\ 5]{thirring1961}; \citealp[box 18.1]{MTW}; \citealp[p.\ 76]{waldGR}; \citealp[sec.\ 2]{straumann2000}).}  It is hard to say what lesson we should draw.  If, despite these objections, you conclude that the field-theoretic approach provides a reasonably good perspective on general relativity, then you can read this section straightforwardly as a discussion of the mass of the gravitational field in general relativity.  If, on the other hand, you see these objections as indicating that the field-theoretic approach is far removed from the standard geometric approach to general relativity, then you can read this section like the two that preceded it: as a discussion of the mass of the gravitational field in a theory of gravity less advanced than true general relativity (though closer to it than either Newtonian gravity or gravitoelectromagnetism) and as a prolegomenon to a real analysis of the mass of the gravitational field in general relativity (which I do not give in this paper).

There a number of challenges facing attempts to define a gravitational energy density in general relativity---that are of course also challenges to defining a mass density for the gravitational field (as mass density is just energy density over $c^2$).  One problem that has been discussed widely in the physics and philosophy literature is that, in the geometric approach to general relativity, the gravitational energy density one assigns to a point depends on one's choice of coordinates.\footnote{See (\citealp[p.\ 307]{landaulifshitzfields}; \citealp[sec.\ 20.4]{MTW}; \citealp[sec.\ 11.2]{waldGR}; \citealp{pintoneto2000, hoefer2000, pitts2010}; \citealp[p.\ 1018]{lam2011}; \citealp{duerrforth}; \citealp{duerr2017, curiel, readforth}).  (\citealp{dewar2018} argue that this problem arises even in Newtonian gravity.)  One common response is to forego a notion of local gravitational energy density in general relativity and to instead seek a way of defining the amount of gravitational energy only globally (through, e.g., the Komar, Bondi, or ADM definitions; \citealp[sec.\ 11.2]{waldGR}) or perhaps defining it only quasi-locally---i.e., in certain finite regions \citep{szabados2009}.\label{GRreferences}}  This problem will not be addressed head-on as it is sidestepped in the field-theoretic approach where the flat background spacetime defines a preferred set of coordinates.  However, there is another widely discussed challenge for defining gravitational energy density in general relativity which cannot be so easily avoided as it afflicts both the geometric and field-theoretic approaches to general relativity: there are multiple ways to define an energy-momentum tensor for the gravitational field.

\subsection{The Conservational Role}\label{secconsroleGR}

It seems clear that we must attribute some mass to the gravitational field in order to ensure conservation of mass.  But, exactly how one ought to do so is less clear.  There are a number of different ways to represent the flow of mass, energy, and momentum in the gravitational field---related to the different ways of describing the density and flow of energy in Newtonian gravity (section \ref{secconsroleNG}).  Here I will discuss two of the existing representations---those of Weinberg and Landau-Lifshitz---and give desiderata for a third.

To understand Weinberg's representation, let us begin by rewriting the Einstein field equations
\begin{equation}
G_{\mu\nu}=R_{\mu\nu}-\frac{1}{2}R g_{\mu\nu}=-\frac{8 \pi G}{c^4}T_{\mu\nu}\ ,
\label{einsteinfieldeq}
\end{equation}
using the decomposition of $g_{\mu\nu}$ into background and field from \eqref{decomposition}:
\begin{equation}
R^{(1)}_{\mu\nu}-\frac{1}{2}R^{(1)}\eta_{\mu\nu}=-\frac{8 \pi G}{c^4} \left[T_{\mu\nu}+t_{\mu\nu}\right]
\ .
\label{sourceGR}
\end{equation}
Here $R^{(1)}_{\mu\nu}$ is the part of the Ricci tensor $R_{\mu\nu}$ which is linear in $h_{\mu\nu}$,
\begin{equation}
R^{(1)}_{\mu\nu}=\frac{1}{2}\left(\partial_\nu \partial_\mu h^\lambda_{\ \lambda} +\partial_\lambda\partial^\lambda h_{\mu\nu}-\partial_\lambda\partial_\nu h^\lambda_{\ \mu}-\partial_\lambda\partial_\mu h^\lambda_{\ \nu}\right)
\ ,
\label{firstorderR}
\end{equation}
and $t_{\mu\nu}$ serves as a catch-all for the higher order terms in $R_{\mu\nu}$,
\begin{align}
t_{\mu\nu}&=\frac{c^4}{8 \pi G}\left(R_{\mu\nu}-\frac{1}{2}R g_{\mu\nu}-R^{(1)}_{\mu\nu}+\frac{1}{2}R^{(1)}\eta_{\mu\nu}\right)\ .
\label{weinbergtensor}
\end{align}
In the equations of general relativity above and below I've adopted the sign conventions and notation of \citep{weinbergGR}, though factors of $c$ are included here.

The tensor $T^{\mu\nu}$ is generally interpreted as the energy-momentum tensor of matter (where here `matter' is meant to include the electromagnetic field and any other non-gravitational fields).  Weinberg calls $t^{\mu\nu}$ `the energy-momentum ``tensor'' of the gravitational field',\footnote{This candidate energy-momentum tensor for the gravitational field also appears in \citep[p.\ 465]{MTW}.} giving a number of reasons why the tensor deserves this name, including the facts that: (a) $T^{\mu\nu}$ and $t^{\mu\nu}$ act together in sum as the source of the gravitational field in \eqref{sourceGR}, and (b) neither the energy-momentum of matter (as described by $T^{\mu\nu}$) nor the energy-momentum of the gravitational field (as described by $t^{\mu\nu}$) are alone conserved, but total energy-momentum is conserved,
\begin{equation}
\frac{\partial}{\partial x^{\mu}}\left[T^{\mu\nu}+t^{\mu\nu}\right]=0
\ .
\label{totalconservation}
\end{equation}
This equation encodes both conservation of energy (and thus also conservation of mass as mass is proportional to energy) and conservation of momentum.  One might contend that the energy-momentum of matter alone is conserved because the covariant four-dimensional divergence of $T^{\mu\nu}$ is zero,
\begin{equation}
T^{\mu\nu}_{\ \ ;\mu}=\frac{\partial}{\partial x^{\mu}}T^{\mu\nu}+\Gamma^{\mu}_{\ \mu\lambda}T^{\lambda\nu}+\Gamma^{\nu}_{\ \mu\lambda}T^{\mu\lambda}=0
\ .
\label{nonconservationlaw}
\end{equation}
However, from the perspective of the field-theoretic approach with $\eta_{\mu\nu}$ as the background spacetime, it is clear that \eqref{totalconservation} is a true conservation law and \eqref{nonconservationlaw} is not.\footnote{Many authors give explanations as to why \eqref{nonconservationlaw} should not be regarded as a conservation law for the energy-momentum of matter alone: \citet[sec.\ 101]{landaulifshitzfields}; \citet[sec.\ 7.6]{weinbergGR}; \citet[sec.\ 1.4.2]{straumann2004}; \citet[sec.\ 8.3]{hobsonGR}; \citet{ohanian2010,lam2011}.  On the other side, \citet[ch.\ 16--17]{MTW} and \citet[pp.\ 69--70]{waldGR} take \eqref{nonconservationlaw} to be a local conservation law for the energy and momentum of matter alone and \citet{dewar2018} argue from this interpretation that (unless one modifies Einstein's field equations) there cannot be an energy assigned to gravity which is ever transferred either to or from matter.}

Both of the reasons given above are fine reasons for taking the total energy-momentum of matter and gravitational field to be given by $\tau^{\mu\nu}=T^{\mu\nu}+t^{\mu\nu}$.  But, it is not so clear that we have properly split the energy-momentum of the gravitational field from the energy-momentum of matter.  We will return to this point shortly.

\citet[sec.\ 101]{landaulifshitzfields} introduce an alternative tensor to describe the flow of energy and momentum in the gravitational field.  Their tensor appears on the righthand side of the Einstein field equations when these equations are written in a third way, different from either \eqref{einsteinfieldeq} or \eqref{sourceGR},
\begin{equation}
\frac{\partial}{\partial x^{\rho}}\frac{\partial}{\partial x^{\sigma}}\left[(-g)(g^{\mu\nu}g^{\rho\sigma}-g^{\mu\rho}g^{\nu\sigma}\right]=\frac{16 \pi G}{c^4}(-g)(T^{\mu\nu}+t_{LL}^{\mu\nu})\ .
\label{einsteinfieldeqLL}
\end{equation}
Here $g$ is the determinant of $g_{\mu\nu}$ and $t_{LL}^{\mu\nu}$ is the Landau-Lifshitz tensor.  The energy and momentum of field and matter together are conserved as,
\begin{equation}
\frac{\partial}{\partial x^{\mu}}\left[(-g)(T^{\mu\nu}+t_{LL}^{\mu\nu})\right]=0
\ .
\label{totalconservation2}
\end{equation}
This is a true conservation law, like \eqref{totalconservation}, provided we revise our understanding of the energy-momentum tensor of matter so that it is given by $-g T^{\mu\nu}$, not $T^{\mu\nu}$, and the energy-momentum tensor for the field is similarly understood as $-g t_{LL}^{\mu\nu}$, not $t_{LL}^{\mu\nu}$.  The Landau-Lifshitz tensor can be defended by considerations similar to the two marshaled in favor of Weinberg's tensor: (a) $T^{\mu\nu}$ and $t_{LL}^{\mu\nu}$ act together in sum as the source of the gravitational field in \eqref{einsteinfieldeqLL}, and (b) the total energy-momentum, $\tau_{LL}^{\mu\nu}=(-g)(T^{\mu\nu}+t_{LL}^{\mu\nu})$, is conserved.  Again, the splitting of $\tau_{LL}^{\mu\nu}$ into matter and field contributions can be questioned.

The objects $t^{\mu\nu}$ and $t_{LL}^{\mu\nu}$ are often called `pseudotensors' because they do not count as true tensors by the lights of the geometric approach to general relativity where $g_{\mu\nu}$ is taken to be the spacetime metric.  However, in the field-theoretic approach adopted here (with a fixed Minkowski spacetime background), $t^{\mu\nu}$ and $t_{LL}^{\mu\nu}$ are genuine tensors because they transform properly under Lorentz transformations.\footnote{See (\citealp[p.\ 306]{landaulifshitzfields}; \citealp[p.\ 167]{weinbergGR}; \citealp{pintoneto2000}; \citealp[sec.\ 1]{pittsschieve2007}).}

In order to better understand the Weinberg and Landau-Lifshitz tensors, let us consider them in the Newtonian limit of general relativity\footnote{Since we are deriving the Newtonian energy formulas, I refer to this approximation as the `Newtonian limit'.  But, as \citet[p.\ 51]{chandrasekhar1969} emphasizes, deriving such expressions requires looking at terms that are normally called `post-Newtonian'.} where $h_{\mu\nu}$ can be written in terms of the gravitational potential $\phi$ as
\begin{equation}
\left(
 \begin{matrix}
  \frac{2 \phi}{c^2} & 0 & 0 & 0 \\
  0 & \frac{2 \phi}{ c^2} & 0 & 0 \\
  0 & 0 & \frac{2 \phi}{c^2} & 0 \\
  0 & 0 & 0 & \frac{2 \phi}{c^2}
 \end{matrix}
\right) + \mbox{ order }\phi^2
\ .
\end{equation}
Here the energy densities are (to lowest order)\footnote{These energy densities are given in (\citealp[eq.\ 46]{chandrasekhar1969}; \citealp[eq.\ 36]{peters1981}; \citealp[eq.\ 57 and 70]{nikishov2001}; \citealp[eq.\ 7.50a]{poissonwill}).}
\begin{align}
t^{00}&=\frac{-3|\del\phi|^2}{8 \pi G}
\nonumber
\\
-g t_{LL}^{00}&=\frac{-7|\del\phi|^2}{8 \pi G}
\ .
\end{align}
These look quite similar to the expressions for the Newtonian energy density in \eqref{energydensityNG3} with $\alpha$ equal to 3 and 7, respectively.  However, the terms proportional to $\rho_m \phi$ are absent.  \citet[p.\ 561]{peters1981} conjectures that $T^{00}$ must include a $-\rho_m \phi$ term so that $T^{00}+t^{00}$ contains the correct gravitational energy in the Newtonian limit, though he does not derive this for any particular type of matter.  \citet[eq.\ 5.1]{kaplan2009} show that, for a perfect fluid, $-gT^{00}$ contains a $-3\rho_m \phi$ term in addition to the non-gravitational energy of matter\footnote{See also \citep[eq.\ 7.47a]{poissonwill}.}---exactly what's needed for $(-g)(T^{00}+t_{LL}^{00})$ to match the energy density in \eqref{energydensityNG3} with $\alpha=7$.\footnote{Here it is Kaplan et al.'s `conserved rest-mass density' $\rho_*$ and not $\rho_0$ which corresponds to the (non-relativistic) mass density $\rho_m$ from section \ref{sectionNG} (see also \citealp[sec.\ 7]{chandrasekhar1965}; \citealp[sec.\ 39.11]{MTW}).}  This term might be interpreted as a potential energy density of matter, an interaction energy density, or a contribution to the field energy density.  If we'd like to interpret it as part of the field energy density (as in section \ref{secconsroleNG}), then we should conclude that it has been misplaced (taking advantage of our freedom to question the splitting of $\tau_{LL}^{\mu\nu}$ into matter and field contributions).  One could redefine the energy-momentum tensors for matter and field so that terms like this are moved from the energy-momentum tensor of matter to the energy-momentum tensor of the field.

In addition to energy density, we can also examine energy flux in the Newtonian limit.  The energy flux density (Poynting vector) derived from the Landau-Lifshitz tensor, $-g c \: t_{LL}^{0i}$, is given in \citep[eq.\ 55]{chandrasekhar1969}.  One can derive\footnote{In this derivation one must make use of equations 3, 45, 49, and 117 in \citep{chandrasekhar1965}, as well as the remark by \citet[sec.\ 3]{kaplan2009} that their $\rho_*$ can be treated as the source of $\phi$.} that the divergence of this Poynting vector is
\begin{equation}
\frac{\partial}{\partial x^{i}}(-g c \: t_{LL}^{0i})=\frac{1}{4\pi G c}\del\cdot\left[ \frac{1}{4\pi G}\left(4\phi  \frac{\partial}{\partial t}  \del \phi + 3\frac{\partial \phi}{\partial t}\del \phi  +4 \vec{v}_m\phi\nabla^2 \phi\right)  \right]
\ ,
\end{equation}
which almost agrees with the divergence of the Newtonian Poynting vector \eqref{energyconsNG3}---as the gravitoelectromagnetic Poynting vector was shown to in section \ref{secconsroleGEM}.  The only difference is the coefficient on the third term.  This is compensated for by a $-3 \vec{v}_m\phi\rho_m$ term which \citet[eq.\ 5.1]{kaplan2009} have shown to appear in the energy flux density of a perfect fluid, $-g c \: T^{0i}$.  Again, one might question whether such a term belongs in the energy-momentum tensor of matter when it could instead be moved into the energy-momentum tensor of the field.

Earlier I mentioned that the field-theoretic approach to general relativity avoids the problem of gravitational energy being coordinate-dependent by introducing a fixed spacetime background.  \citet[p.\ 700]{pittsschieve2007}\footnote{See also (\citealp{pintoneto2000}; \citealp{pitts2010}).} claim that, in avoiding this problem, the field-theoretic approach ends up facing another: the tensors that one might introduce to describe the flow of energy and momentum in the gravitational field---like those of Weinberg and Landau-Lifshitz---are gauge-dependent.  This is what one would expect from their limiting behavior.  The general Newtonian energy density \eqref{energydensityNG3} is only gauge-independent for $\alpha=1$ (Maxwell's proposal) and these tensors correspond to $\alpha=3$ and $7$.  A gravitational energy-momentum tensor which limits to \eqref{energydensityNG3} for $\alpha=1$ might fare better regarding gauge-dependence.

Many authors have lamented the proliferation of energy-momentum tensors for the gravitational field.  But, I think there is good reason to search for another.  In each of the three theories we considered previously, one was able to fully separate the energy and momentum of the field from the energy and momentum of matter so that no field-dependent quantities appeared in the energy and momentum densities (or flux densities) for matter.  By recarving the total energy-momentum tensor into field and matter contributions, one should be able to do so in general relativity as well.  Beyond this freedom to recarve, there is a second freedom available in seeking new stress-energy tensors: one can rewrite the Einstein field equations so that different terms appear on the righthand side, as in \eqref{sourceGR} and \eqref{einsteinfieldeqLL}.  We've already seen that such rewritings correspond to different Newtonian energy densities in the Newtonian limit.  It would be worthwhile to use these two freedoms to find a rewriting and recarving for which the energy-momentum tensor of the field yields Maxwell's Newtonian energy density \eqref{energydensityNG} and the corresponding energy flux density \eqref{PoyntingNG} in the Newtonian limit---as well as the gravitoelectromagnetic energy density \eqref{energydensityGEM} and energy flux density \eqref{PoyntingGEM} on the way to that limit.  This would bring our understanding of energy and momentum in general relativity into closer alignment with these two less-advanced theories, and also with electromagnetism (as it is so similar to gravitoelectromagnetism).

\subsection{The Inertial Role}

The conservation laws above---\eqref{totalconservation} and \eqref{totalconservation2}---contain conservation of momentum equations similar to those examined earlier---\eqref{forcelawEM}, \eqref{forcelawNG}, and \eqref{forcelawGEM}---with the field velocity given by either $c \: t^{0i}/t^{00}$ or $c \: t_{LL}^{0i}/t_{LL}^{00}$.  In these equations, as in the similar ones considered before, the mass of the field plays the inertial role.

As was done before, we can also examine the way in which the mass of the gravitational field plays the inertial role indirectly by asking whether it requires more force to accelerate a massive body with gravity turned on or gravity turned off.  In section \ref{secinertialroleEM} we identified two effects modifying the reaction of a charged body to applied forces: (a) the mass of the electromagnetic field surrounding the body must be accelerated, and (b) electromagnetic waves are produced when the body is accelerated and these waves carry away energy and momentum.  In section \ref{secinertialroleGEM} we saw that in gravitoelectromagnetism massive bodies are surrounded by clouds of negative gravitational field mass which make those bodies easier to accelerate.  When such bodies are accelerated they emit gravitational waves which carry away negative energy.  In general relativity, as in gravitoelectromagnetism, the total mass of the gravitational field around a body is negative---at least when the gravitational energy density is well-approximated by the Newtonian expression \eqref{energydensityNG3} for some value of $\alpha$.  Again, this negative field mass makes the body easier to accelerate.  Unlike gravitoelectromagnetism, the energy of gravitational waves in general relativity is positive.  This makes a body harder to accelerate from rest---in opposition to the first effect---since one must also supply the energy that goes into the gravitational waves.\footnote{\citet[ch.\ 12]{poissonwill} analyze this gravitational radiation reaction in detail and compare it to the (much simpler) case of electromagnetic radiation reaction (see footnote \ref{emradreaction}).}

It is worth remembering that these two effects are generally small.  The mass in the gravitational field around the earth is ten orders of magnitude smaller than the mass of the earth itself and the earth emits an insignificant amount of gravitational radiation as it orbits the sun.  However, these effects can be significant.  Gravitational waves were first observed indirectly by their radiation reaction effects.  The orbits of binary pulsars were seen to change over time as the pulsars lost energy to gravitational waves.\footnote{See \citep[sec.\ 12.4.1]{poissonwill}, cf. \citep{duerr2017}.}  Gravitational waves have now been directly observed by the Laser Interferometer Gravitational-Wave Observatory (LIGO).  The first gravitational waves observed by LIGO came from a black hole merger in which so much energy was radiated away through gravitational waves that after the black holes had merged their total mass had decreased by about three times the mass of the sun \citep{abbott2016}.

\subsection{The Gravitational Role}\label{secgravroleGR}

Looking at the Einstein field equations \eqref{einsteinfieldeq}, it appears that the only source for gravity is the energy-momentum of matter, $T^{\mu\nu}$, on the righthand side and thus that gravitational energy-momentum is not a source of gravity.  \citet[p.\ 467]{MTW} write that,
\begin{quote}
...`local gravitational energy-momentum' has no weight.  It does not curve space.  It does not serve as a source term on the righthand side of Einstein's field equations.  It does not produce any relative geodesic deviation of two nearby world lines that pass through the region of space in question.  It is not observable.\footnote{\citet[p.\ 467]{MTW} write that the gravitational energy in a region can act as a source of gravitation (even though locally gravitational energy does not act as a source).  They also acknowledge (pp.\ 424--425, 437) that, in the field-theoretic approach, the local energy-momentum of the gravitational field is treated as a source for the gravitational field.}
\end{quote}
However, there are a couple of reasons to doubt this assessment.  First, as was discussed earlier, it may be the case that some of the terms that appear in $T^{\mu\nu}$ are best interpreted as describing the energy and momentum of the gravitational field (such terms having been improperly lumped into what ordinarily gets called the energy-momentum tensor for matter).  If that's correct, then when Einstein's field equations are written in the standard from \eqref{einsteinfieldeq} there actually are source terms corresponding to the energy and momentum of the gravitational field appearing on the righthand side (though they're hidden in $T^{\mu\nu}$).\footnote{\citet{lehmkuhl2011} discusses the dependence of $T^{\mu\nu}$ on the metric field and considers how this complicates its role as source for gravitation (see especially footnote 49).}  Second, it is possible to move some terms in \eqref{einsteinfieldeq} from the left to the right and to take the righthand side of, for example, \eqref{sourceGR} or \eqref{einsteinfieldeqLL} instead to be the source of gravity.  In each of these equations, an energy-momentum tensor for the gravitational field appears as a source on the righthand side.  \citet[p.\ 151]{weinbergGR} writes the Einstein field equations in one of these alternate ways \eqref{sourceGR} with the linear terms in $h_{\mu\nu}$ appearing on the lefthand side (characterizing the reaction of the field to sources) and the nonlinear terms appearing on the right (interpreted as source terms).  He clearly states his view that gravity acts as a source for itself in explaining why general relativity is so much more complicated than electromagnetism:
\begin{quote}
Maxwell's equations are linear because the electromagnetic field does not itself carry charge, whereas gravitational fields do carry energy and momentum ... and must therefore contribute to their own source.  That is, the gravitational field equations will have to be nonlinear partial differential equations, the nonlinearity representing the effect of gravitation on itself.
\end{quote}
What exactly it is that acts as source for gravitation appears to depend on the way we choose to write the Einstein field equations.  This is because the idea of a `source' is only really defined relative to a mathematical expression for the reaction of something to that putative source; for example, by the lefthand side of \eqref{einsteinfieldeq}, \eqref{sourceGR}, or \eqref{einsteinfieldeqLL}.

In general relativity as in gravitoelectromagnetism, massive bodies are surrounded by clouds of gravitational field mass.  How exactly this field mass is distributed will depend on which energy-momentum tensor is used for the field.  For the two tensors considered above in the Newtonian limit, the energy densities are given by \eqref{energydensityNG3} with $\alpha$ equal to either 3 or 7.  For these values of alpha, the energy density is the sum of one negative term which extends beyond the bounds of the massive body as well as a positive energy density which is confined to the region where the mass density is nonzero.  If we find an energy-momentum tensor which yields a Newtonian field energy density with $\alpha=1$ (as was wished for in section \ref{secconsroleGR}), then the field energy in this limit will be purely negative.  One might hope to decide between such energy densities by looking at the gravitational pull of the gravitational field's mass \citep{peters1981}.  Where the field's energy is will determine where it acts as a source.  Unfortunately, the laws that describe how the gravitational field acts as a source for itself---like \eqref{sourceGR} and \eqref{einsteinfieldeqLL}---covary with the different energy-momentum tensors for the field so that wherever the field's mass is, it is regarded as acting as a source there (by the lights of whichever version of the Einstein field equations is appropriate to it).

\section{Conclusion}

Let us review whether the mass of the gravitational field plays each of the three roles we've been discussing in the three different theories of gravity that have been considered: Newtonian gravity, gravitoelectromagnetism, and general relativity (in its nonstandard field-theoretic formulation).

In general relativity and in gravitoelectromagnetism, the mass of the gravitational field plays the conservational role.  As energy is exchanged between matter and the gravitational field, the relativistic mass of matter does not remain constant.  However, the total mass of matter and field does.  In the non-relativistic theory of Newtonian gravity, the mass of matter does not change as it exchanges energy with the gravitational field.  The gravitational field has no mass playing the conservational role.  In general relativity and Newtonian gravity, we noted that there are multiple energy densities one could assign to the gravitational field---and that the different general relativistic energy densities correspond to different Newtonian energy densities.

In general relativity and in gravitoelectromagnetism, the mass of the gravitational field plays the inertial role.  One can write an Eulerian force law describing the reaction of the field to forces from matter acting upon it and in this law the mass of the field quantifies the resistance to acceleration of the field itself.  In Newtonian gravity one can write a similar equation, but in that equation the field has no momentum because it has no mass.  In general relativity and in gravitoelectromagnetism, massive bodies are made easier to accelerate by the negative energy gravitational fields surrounding them.  When such a body is accelerated it loses energy by emitting gravitational radiation in general relativity and gains energy from emitting radiation in gravitoelectromagnetism.

In general relativity, the gravitational field appears to act as a source for itself, but the way it does so is not entirely clear as there is ambiguity in what it means for something to act as a source for itself.  In gravitoelectromagnetism, the close analogy with electromagnetism makes it easier to see whether gravity acts as a source for itself.  To avoid inconsistency, gravity must act as a source for itself.  In Newtonian gravity the only source of the gravitational field is matter.  The field has no mass playing the gravitational role.\\

\noindent
\textbf{Acknowledgments}
Thank you to Craig Callender, Erik Curiel, Neil Dewar, John Dougherty, Dennis Lehmkuhl, Jim Weatherall, and anonymous referees for helpful feedback and discussion.  This project was supported in part by funding from the President's Research Fellowships in the Humanities, University of California (for research conducted while at the University of California, San Diego).  I would also like to thank LMU Munich for their hospitality during a visit in the winter of 2018.

\appendix

\section{Mass-Energy Equivalence}\label{MassEnergyEquivSec}

There has been some debate as to how one should properly understand mass-energy equivalence in the context of relativistic theories.  As I understand it, mass-energy equivalence is the claim that anything which has relativistic mass $m_r$ has energy $m_r c^2$ and anything with energy $\mathcal{E}$ has relativistic mass $\mathcal{E}/c^2$.\footnote{This way of understanding mass-energy equivalence follows (\citealp[sec.\ 34]{fock1959}; \citealp{bondispurgin}; \citealp[sec.\ 6.3]{rindler2006}).}  This is meant to apply to all things that possess energy: particles, fields, systems of particle and field, fluids, etc.  A special case of this equivalence is that in something's rest frame, its energy is equal to its proper mass $m_0$ times $c^2$ (as $m_r=m_0$ in the rest frame).  The equation $\mathcal{E}=m_r c^2$ is universally valid, but $\mathcal{E}=m_0 c^2$ is only valid in the rest frame.\footnote{This point has been emphasized by many authors, including \citeauthor{lange2001} (\citeyear{lange2001}, \citeyear{lange}); \citet[sec.\ 1.1]{fernflores}.}

Because relativistic mass and energy are directly proportional to one another and cannot vary independently, I don't think it is unreasonable to say that: although relativistic mass and energy `are measured by different units', they are `two ways of measuring what is essentially the same thing' \citep[p.\ 146]{eddington}.\footnote{See also (\citealp[p.\ 197]{einsteininfeld}; \citealp[ch.\ 13]{jammer1961}; \citealp[p.\ 225]{lange2001}; \citealp[p.\ 239]{lange}; \citealp[sec.\ 2.5]{fernflores}; \citealp[ch.\ 3]{fernfloresvol2}).}  Noting this tight connection, you can read this paper as a discussion of the extent to which field energy plays the traditional roles of mass (in different classical field theories).\footnote{In this spirit, Rindler writes that: `The formula $E=mc^2$ [where $m$ is relativistic mass] reminds me that energy has mass-like properties such as inertia and gravity...' \citep{rindler1990}.}

As I've formulated mass-energy equivalence, it is not fundamentally a claim about processes that, in some sense, involve the conversion of mass into energy (or vice versa).  Mass-energy equivalence is first and foremost about the properties things have, not about the way those properties evolve over time.  However, because energy is conserved in every frame, the principle constrains what can happen in processes that have been described as conversions of mass into energy.\footnote{By mass-energy equivalence, conservation of relativistic mass follows immediately from conservation of energy.  If there exists a frame in which the system is at rest, conservation of proper mass follows immediately from conservation of relativistic mass (as the proper mass and relativistic mass of the system are equal in that frame).}  If an isolated system begins some process with a certain energy, relativistic mass, and proper mass, it must end the process with the same energy, relativistic mass, and proper mass.\footnote{The fact that---for an isolated system---energy, relativistic mass, and proper mass are each conserved is discussed clearly by \citeauthor{lange2001} (\citeyear{lange2001}, \citeyear{lange}) and \citet{fernflores}, using examples like \eqref{reactionexample}.  \citet{bondispurgin} correctly highlight the conservation of energy and relativistic mass, though they confusingly write that `rest mass is generally not conserved' because the sum of the rest masses of a pre-interaction collection of bodies may be less than sum of the rest masses of the bodies that are present after the interaction has been completed---as occurs in \eqref{reactionexample}.}  For example, consider colliding an electron and positron to produce a pair of photons (particles that have no proper mass),
\begin{equation}
e^{-}+e^{+} \longrightarrow \gamma + \gamma
\ .
\label{reactionexample}
\end{equation}
Viewed from the rest frame of the incoming electron, the energy $\mathcal{E}=m_0 c^2$ of the electron (associated with its proper mass) is converted into kinetic energy of the two photons (as is the energy of the positron).  Strictly speaking, this is not a conversion of the electron's mass into energy, but a conversion of one kind of energy into another.  If the masses of the electron and positron were truly converted into energy, you'd expect the mass of the system to decrease and the energy to increase as a result of the interaction.  But, this is not what happens.  The energy, relativistic mass, and proper mass of the system remain constant.  After the interaction, it may seem like the system has lost proper mass because the two photons each have zero proper mass.  But, the proper mass of the system is not the sum of the proper masses of its parts (unlike relativistic mass, proper mass is not additive).  The two photons that are emitted together have non-zero proper mass because they have non-zero energy in their rest frame (the frame in which the photons have equal and opposite momenta).

\section{Deriving Gravitoelectromagnetism from General Relativity}\label{GEMfromGR}

To derive the equations of gravitoelectromagnetism, let us begin by taking a linear approximation to general relativity (assuming that $h_{\mu\nu}$ in \eqref{decomposition} is sufficiently small that terms second-order in it can be neglected).  In that case, we can write the Einstein field equations \eqref{einsteinfieldeq} as,
\begin{equation}
\partial_\nu\partial_\mu h^\lambda_{\ \lambda} + \partial_\lambda\partial^\lambda h_{\mu\nu}-\partial_\nu\partial_\rho h^\rho_{\ \mu} - \eta_{\mu\nu}(\partial_\lambda\partial^\lambda h-\partial_\rho\partial_\sigma h^{\sigma\rho})=-\frac{16 \pi G}{c^4}T_{\mu\nu} 
\end{equation}
\citep[eq.\ 7.6.19]{weinbergGR}.  Or, in terms of the trace-reverse of $h_{\mu\nu}$,
\begin{equation}
\bar{h}_{\mu\nu}=h_{\mu\nu}-\frac{1}{2}\eta_{\mu\nu}h^\lambda_{\ \lambda}\ ,
\end{equation}
the field equations can be written as
\begin{equation}
\partial_\lambda\partial^\lambda \bar{h}_{\mu\nu}+\eta_{\mu\nu}\partial_\rho\partial_\sigma \bar{h}^{\rho\sigma}-\partial_\nu\partial_\rho \bar{h}^\rho_{\ \mu}-\partial_\mu\partial_\rho \bar{h}^\rho_{\ \nu}=-\frac{16 \pi G}{c^4}T_{\mu\nu} 
\ .
\end{equation}
If we choose the Lorenz gauge\footnote{For an explanation of the gauge freedom you have in general relativity, see (\citealp[boxes 7.1 and 18.2]{MTW}; \citealp[sec.\ 3.7]{feynmanGR}; \citealp[sec.\ 3.2 and 7.1]{ohanianGR}).} for $\bar{h}_{\mu\nu}$ in which $\partial_\mu \bar{h}^{\mu\nu}=0$, the Einstein field equations simplify to
\begin{equation}
\partial_\lambda\partial^\lambda \bar{h}_{\mu\nu}=-\frac{16 \pi G}{c^4}T_{\mu\nu} \ .
\label{simpleeinstein}
\end{equation}
If the $T_{ij}$ terms are sufficiently small and there is no source-free gravitational radiation, the $\bar{h}_{ij}$ terms are negligible and only $\bar{h}_{0\mu}$ needs to be considered.  Let us introduce a gravitational four-potential, defined in terms of $\bar{h}_{0\mu}$ via $A_g^{\mu}=\frac{c^2}{4 \sqrt{G}}\bar{h}^{0\mu}$.  The zeroth component of the Lorenz gauge condition above yields the familiar Lorenz gauge condition of electromagnetism,
\begin{equation}
\frac{1}{c}\frac{\partial A_g^0}{\partial t} = \vec{\nabla}\cdot\vec{A}_g
\end{equation}

The $0\mu$ component of \eqref{simpleeinstein} is
\begin{equation}
\partial_\lambda\partial^\lambda A_g^\mu = \frac{4 \pi \sqrt{G}}{c^2}T^{0\mu}
\ .
\label{maxwellshortform}
\end{equation}
This has exactly the same form as the electromagnetic Maxwell equations for the four-potential in the Lorenz gauge,\footnote{See \citep[eq.\ 3.58b]{MTW}.} except that four-current has been replaced by four-momentum.  The two gravitoelectromagnetic Maxwell equations with sources are just \eqref{maxwellshortform} expressed in terms of the gravitoelectric and gravitomagnetic fields---related to the four-potential by $\vec{E}_g=-\vec{\nabla}A_g^0 - \frac{1}{c}\frac{\partial \vec{A}_g}{\partial t}$ and $\vec{B}_g=\vec{\nabla}\times\vec{A}_g$.  The sourceless gravitoelectromagnetic Maxwell equations follow automatically from the way $\vec{E}_g$ and $\vec{B}_g$ are defined in terms of a four-potential.

Having derived the gravitoelectromagnetic Maxwell equations, we must still derive the gravitoelectromagnetic Lorentz force law.  To do so, let's begin with the geodesic equation of general relativity,
\begin{align}
\frac{d^2 x^i}{d\tau^2}&=-\Gamma^{i}_{\ \alpha\beta}\frac{d x^\alpha}{d\tau}\frac{d x^\beta}{d\tau}
\ ,
\end{align}
and assume that that the velocity of our test particle is small enough that we can ignore terms of order $\frac{v^2}{c^2}$:
\begin{align}
\vec{a}=\frac{d^2 x^i}{dt^2}&=-c^2\Gamma^{i}_{\ 00}-c\Gamma^{i}_{\ j0}\frac{d x^j}{dt}-c\Gamma^{i}_{\ 0j}\frac{d x^j}{dt}
\ .
\end{align}
Here
\begin{align}
\Gamma^{i}_{\ j0}=\Gamma_{ij0}=\Gamma_{i0j}&=\frac{1}{2}(\partial_j h_{0i} - \partial_i h_{j0})
\nonumber
\\
&=\frac{1}{2}(\partial_j \bar{h}_{0i} - \partial_i \bar{h}_{j0})
\nonumber
\\
&=\frac{2\sqrt{G}}{c}(\partial_j A_{g_i} - \partial_i A_{g_j})
\nonumber
\\
&=\frac{-2\sqrt{G}}{c}\epsilon_{ijk}B_{g_k}\ .
\end{align}
Inserting this connection, the geodesic equation becomes
\begin{align}
\vec{a}&=\frac{c^2}{2}\partial_i \bar{h}_{00}+\frac{c^2}{4}\partial_i \bar{h}_{00}+\frac{4\sqrt{G}}{c}\epsilon_{ijk}\frac{d x^j}{dt} B_{g_k}
\nonumber
\\
&=\sqrt{G}\left(\vec{E}_g+\frac{4}{c}\vec{v}\times\vec{B}_g\right)
\ .
\end{align}
Multiplying by the proper mass of the test particle on both sides, this gives a force law
\begin{equation}
\vec{f}_g=m_0\vec{a}=\sqrt{G}m_0\left(\vec{E}_g+\frac{4}{c}\vec{v}\times\vec{B}_g\right)\ ,
\end{equation}
which agrees with \eqref{lorentzforcelawGEM} (though it is applied to a particle instead of a continuum) in the low velocity limit where proper mass is equal to relativistic mass.


\begin{thebibliography}{101}
\expandafter\ifx\csname natexlab\endcsname\relax\def\natexlab#1{#1}\fi
\expandafter\ifx\csname url\endcsname\relax
  \def\url#1{{\tt #1}}\fi
\expandafter\ifx\csname urlprefix\endcsname\relax\def\urlprefix{URL }\fi

\bibitem[{Abbott({[2016]})}]{abbott2016}
Abbott, B. P. \emph{et al.} {[2016]}:  `Observation of Gravitational Waves from a Binary
  Black Hole Merger',
\newblock {\em Physical Review Letters\/}, {\textbf{ 116}\/}(6), pp. 061102.

\bibitem[{Arnowitt \emph{et~al.}({[1960]})Arnowitt, Deser, and
  Misner}]{arnowitt1960}
Arnowitt, R., Deser, S.  and Misner, C.~W. {[1960]}:  `Note on
  Positive-Definiteness of the Energy of the Gravitational Field',
\newblock {\em Annals of Physics\/}, {\textbf{ 11}\/}(1), pp. 116--121.

\bibitem[{Arora and Geppert({[1967]})}]{arora1967}
Arora, R.~K.  and Geppert, D.~V. {[1967]}:  `Comments on ``Energy-Transport
  Velocity in Electromagnetic Waves'' and ``A Classical Electromagnetic Wave
  Paradox''',
\newblock {\em Proceedings of the IEEE\/}, {\textbf{ 55}\/}(3), pp. 452--453.

\bibitem[{Baker({[unpublished]})}]{baker2018}
Baker, D.~J. {[unpublished]}:  `On Spacetime Functionalism',
\newblock available at
  $\langle$\href{http://philsci-archive.pitt.edu/14301/}{philsci-archive.pitt.edu/14301/}$\rangle$.

\bibitem[{Bakopoulos and Kanti({[2017]})}]{bakopoulos2017}
Bakopoulos, A.  and Kanti, P. {[2017]}:  `Novel Ansatzes and Scalar Quantities
  in Gravito-Electromagnetism',
\newblock {\em General Relativity and Gravitation\/}, {\textbf{ 49}\/}, pp. 44.

\bibitem[{Bondi({[1962]})}]{bondi1962}
Bondi, H. {[1962]}:  `On the Physical Characteristics of Gravitational Waves',
\newblock in M.~A. Lichnerowicz  and M.~A. Tonnelat (\emph{eds}), {\em Les
  Th\'{e}ories Relativistes de la Gravitation\/}, \'{E}ditions du Centre
  National de la Recherche Scientifique, pp. 129--135.

\bibitem[{Bondi and Spurgin({[1987]})}]{bondispurgin}
Bondi, H.  and Spurgin, C.~B. {[1987]}:  `Energy has Mass: A Common
  Misunderstanding is Re-examined',
\newblock {\em Physics Bulletin\/}, {\textbf{ 38}\/}(2), pp. 62--63.

\bibitem[{Born and Wolf({[1970]})}]{bornwolf}
Born, M.  and Wolf, E. {[1970]}: {\em Principles of Optics\/},
\newblock Pergamon Press, 4th edition.

\bibitem[{Braginsky \emph{et~al.}({[1977]})Braginsky, Caves, and
  Thorne}]{braginsky1977}
Braginsky, V.~B., Caves, C.~M.  and Thorne, K.~S. {[1977]}:  `Laboratory
  Experiments to Test Relativistic Gravity',
\newblock {\em Physical Review D\/}, {\textbf{ 15}\/}, pp. 2047--2068.

\bibitem[{Chandrasekhar({[1965]})}]{chandrasekhar1965}
Chandrasekhar, S. {[1965]}:  `The Post-Newtonian Equations of Hydrodynamics in
  General Relativity',
\newblock {\em Astrophysical Journal\/}, {\textbf{ 142}\/}, pp. 1488--1512.

\bibitem[{Chandrasekhar({[1969]})}]{chandrasekhar1969}
Chandrasekhar, S. {[1969]}:  `Conservation Laws in General Relativity and in
  the Post-Newtonian Approximations',
\newblock {\em Astrophysical Journal\/}, {\textbf{ 158}\/}, pp. 45--54.

\bibitem[{Ciufolini and Wheeler({[1995]})}]{ciufolini1995}
Ciufolini, I.  and Wheeler, J.~A. {[1995]}: {\em Gravitation and Inertia\/},
\newblock Princeton University Press.

\bibitem[{Clark and Tucker({[2000]})}]{clark2000}
Clark, S.~J.  and Tucker, R.~W. {[2000]}:  `Gauge Symmetry and
  Gravito-Electromagnetism',
\newblock {\em Classical and Quantum Gravity\/}, {\textbf{ 17}\/}(19), pp.
  4125--4157.

\bibitem[{Curiel({[forthcoming]})}]{curiel}
Curiel, E. {[forthcoming]}:  `On Geometric Objects, the Non-Existence of a
  Gravitational Stress-Energy Tensor, and the Uniqueness of the Einstein Field
  Equation',
\newblock {\em Studies in History and Philosophy of Modern Physics\/}.

\bibitem[{Damour \emph{et~al.}({[1991]})Damour, Soffel, and Xu}]{DSX1991}
Damour, T., Soffel, M.  and Xu, C. {[1991]}:  `General-Relativistic Celestial
  Mechanics. I. Method and Definition of Reference Systems',
\newblock {\em Physical Review D\/}, {\textbf{ 43}\/}, pp. 3273--3307.

\bibitem[{Deser({[1970]})}]{deser1970}
Deser, S. {[1970]}:  `Self-Interaction and Gauge Invariance',
\newblock {\em General Relativity and Gravitation\/}, {\textbf{ 1}\/}, pp.
  9--18.

\bibitem[{Dewar and Weatherall({[2018]})}]{dewar2018}
Dewar, N.  and Weatherall, J.~O. {[2018]}:  `On Gravitational Energy in
  Newtonian Theories',
\newblock {\em Foundations of Physics\/}, {\textbf{ 48}\/}(5), pp. 558--578.

\bibitem[{D\"{u}rr({[forthcoming a]})}]{duerrforth}
D\"{u}rr, P.~M. {[forthcoming a]}:  `Fantastic Beasts and Where (not) to Find
  Them: Local Gravitational Energy and Energy Conservation in General
  Relativity',
\newblock {\em Studies in History and Philosophy of Modern Physics\/}.

\bibitem[{D\"{u}rr({[forthcoming b]})}]{duerr2017}
D\"{u}rr, P.~M. {[forthcoming b]}:  `It Ain't Necessarily So: Gravitational
  Waves and Energy Transport',
\newblock {\em Studies in History and Philosophy of Modern Physics\/}.

\bibitem[{Earman and Norton({[1987]})}]{earman1987}
Earman, J.  and Norton, J. {[1987]}:  `What Price Spacetime Substantivalism?
  The Hole Story',
\newblock {\em The British Journal for the Philosophy of Science\/}, {\textbf{
  38}\/}(4), pp. 515--525.

\bibitem[{Eddington({[1920]})}]{eddington}
Eddington, A. {[1920]}: {\em Space, Time and Gravitation: An Outline of the
  General Relativity Theory\/},
\newblock Cambridge University Press.

\bibitem[{Einstein({[1912]})}]{einstein1912}
Einstein, A. {[1912]}:  `Zur Theorie des Statschen Gravitationsfeldes (On the
  Theory of the Static Gravitational Field)',
\newblock {\em Annalen der Physik\/}, {\textbf{ 343}\/}, pp. 443--458,\
\newblock translation in pp.\ 107--120 of \citep{collectedpapers4}.

\bibitem[{Einstein({[1996]})}]{collectedpapers4}
Einstein, A. {[1996]}: {\em The Collected Papers of Albert Einstein, Volume 4:
  The Swiss Years: Writings 1912-1914 (English translation supplement)\/},
\newblock Princeton University Press,\
\newblock translated by Anna Beck.

\bibitem[{Einstein and Infeld({[1961]})}]{einsteininfeld}
Einstein, A.  and Infeld, L. {[1961]}: {\em The Evolution of Physics: From
  Early Concepts to Relativity and Quanta\/},
\newblock Cambridge University Press, 2nd edition.

\bibitem[{Fernflores({[2012]})}]{fernflores}
Fernflores, F. {[2012]}:  `The Equivalence of Mass and Energy',
\newblock in E.~N. Zalta (\emph{ed.}), {\em The Stanford Encyclopedia of
  Philosophy,\/}
\newblock available at
  $\langle$\href{https://plato.stanford.edu/archives/spr2012/entries/equivME/}{plato.stanford.edu/archives/spr2012/entries/equivME/}$\rangle$.

\bibitem[{Fernflores({[2018]})}]{fernfloresvol2}
Fernflores, F. {[2018]}: {\em Einstein's Mass-Energy Equation, Volume II:
  Quantum Mechanics and Gravitation, Empirical Tests, and Philosophical
  Debates\/},
\newblock Momentum Press.

\bibitem[{Feynman \emph{et~al.}({[1995]})Feynman, Morinigo, and
  Wagner}]{feynmanGR}
Feynman, R.~P., Morinigo, F.~B.  and Wagner, W.~G. {[1995]}: {\em Feynman
  Lectures on Gravitation\/},
\newblock B. Hatfield (\emph{ed.}), Addison-Wesley,\
\newblock foreword by John Preskill and Kip S. Thorne.

\bibitem[{Fock({[1959]})}]{fock1959}
Fock, V. {[1959]}: {\em The Theory of Space Time and Gravitation\/},
\newblock Pergamon Press,\
\newblock translated by N. Kemmer.

\bibitem[{Forward({[1961]})}]{forward1961}
Forward, R. {[1961]}:  `General Relativity for the Experimentalist',
\newblock {\em Proceedings of the IRE\/}, {\textbf{ 49}\/}(5), pp. 892--904.

\bibitem[{Franklin({[2015]})}]{franklin2015}
Franklin, J. {[2015]}:  `Self-Consistent, Self-Coupled Scalar Gravity',
\newblock {\em American Journal of Physics\/}, {\textbf{ 83}\/}(4), pp.
  332--337.

\bibitem[{Frisch({[2005]})}]{frisch2005}
Frisch, M. {[2005]}: {\em Inconsistency, Asymmetry, and Non-Locality: A
  Philosophical Investigation of Classical Electrodynamics\/},
\newblock Oxford University Press.

\bibitem[{Geppert({[1965]})}]{geppert1965}
Geppert, D.~V. {[1965]}:  `Energy-Transport Velocity in Electromagnetic Waves',
\newblock {\em Proceedings of the IEEE\/}, {\textbf{ 53}\/}(11), pp. 1790.

\bibitem[{Geroch({[1978]})}]{geroch1978}
Geroch, R. {[1978]}:  `The Positive-Mass Conjecture',
\newblock in N.~Lebovitz, W.~Reid  and P.~Vandervoort (\emph{eds}), {\em
  Theoretical Principles in Astrophysics and Relativity\/}, University of
  Chicago Press, pp. 245--252.

\bibitem[{Giulini({[1997]})}]{giulini1997}
Giulini, D. {[1997]}:  `Consistently Implementing the Field Self-Energy in
  Newtonian Gravity',
\newblock {\em Physics Letters A\/}, {\textbf{ 232}\/}(3), pp. 165--170.

\bibitem[{Griffiths({[1999]})}]{griffiths}
Griffiths, D.~J. {[1999]}: {\em Introduction to Electrodynamics\/},
\newblock Prentice Hall, 3rd edition.

\bibitem[{Gupta({[1957]})}]{gupta1957}
Gupta, S.~N. {[1957]}:  `Einstein's and Other Theories of Gravitation',
\newblock {\em Reviews of Modern Physics\/}, {\textbf{ 29}\/}(3), pp. 334--336.

\bibitem[{Harris({[1991]})}]{harris1991}
Harris, E.~G. {[1991]}:  `Analogy Between General Relativity and
  Electromagnetism for Slowly Moving Particles in Weak Gravitational Fields',
\newblock {\em American Journal of Physics\/}, {\textbf{ 59}\/}(5), pp.
  421--425.

\bibitem[{Heaviside({[1893]})}]{heaviside1893}
Heaviside, O. {[1893]}:  `A Gravitational and Electromagnetic Analogy, Parts I
  and II',
\newblock {\em The Electrician\/}, {\textbf{ 31}\/}, pp. 281--282, 359,\
\newblock reprinted in \citep{jefimenko2000}.

\bibitem[{Hobson \emph{et~al.}({[2006]})Hobson, Efstathiou, and
  Lasenby}]{hobsonGR}
Hobson, M.~P., Efstathiou, G.~P.  and Lasenby, A.~N. {[2006]}: {\em General
  Relativity: An Introduction for Physicists\/},
\newblock Cambridge University Press.

\bibitem[{Hoefer({[1996]})}]{hoefer1996}
Hoefer, C. {[1996]}:  `The Metaphysics of Space-Time Substantivalism',
\newblock {\em The Journal of Philosophy\/}, {\textbf{ 93}\/}(1), pp. 5--27.

\bibitem[{Hoefer({[2000]})}]{hoefer2000}
Hoefer, C. {[2000]}:  `Energy Conservation in GTR',
\newblock {\em Studies in History and Philosophy of Modern Physics\/},
  {\textbf{ 31}\/}(2), pp. 187--199.

\bibitem[{Jackson({[1999]})}]{jackson}
Jackson, J.~D. {[1999]}: {\em Classical Electrodynamics\/},
\newblock Wiley, 3rd edition.

\bibitem[{Jammer({[1961]})}]{jammer1961}
Jammer, M. {[1961]}: {\em Concepts of Mass in Classical and Modern Physics\/},
\newblock Harvard University Press.

\bibitem[{Jantzen \emph{et~al.}({[1992]})Jantzen, Carini, and
  Bini}]{jantzen1992}
Jantzen, R.~T., Carini, P.  and Bini, D. {[1992]}:  `The Many Faces of
  Gravitoelectromagnetism',
\newblock {\em Annals of Physics\/}, {\textbf{ 215}\/}(1), pp. 1--50.

\bibitem[{Jefimenko({[2000]})}]{jefimenko2000}
Jefimenko, O.~D. {[2000]}: {\em Causality, Electromagnetic Induction, and
  Gravitation\/},
\newblock Electret Scientific Company.

\bibitem[{Jefimenko({[2006]})}]{jefimenko2006}
Jefimenko, O.~D. {[2006]}: {\em Gravitation and Cogravitation\/},
\newblock Electret Scientific Company.

\bibitem[{Kaplan \emph{et~al.}({[2009]})Kaplan, Nichols, and
  Thorne}]{kaplan2009}
Kaplan, J.~D., Nichols, D.~A.  and Thorne, K.~S. {[2009]}:  `Post-Newtonian
  Approximation in Maxwell-Like Form',
\newblock {\em Physical Review D\/}, {\textbf{ 80}\/}, pp. 124014.

\bibitem[{Keppel \emph{et~al.}({[2009]})Keppel, Nichols, Chen, and
  Thorne}]{keppel2009}
Keppel, D., Nichols, D.~A., Chen, Y.  and Thorne, K.~S. {[2009]}:  `Momentum
  Flow in Black-Hole Binaries. I. Post-Newtonian Analysis of the Inspiral and
  Spin-Induced Bobbing',
\newblock {\em Physical Review D\/}, {\textbf{ 80}\/}, pp. 124015.

\bibitem[{Knox({[forthcoming]})}]{knoxF}
Knox, E. {[forthcoming]}:  `Physical Relativity from a Functionalist
  Perspective',
\newblock {\em Studies in History and Philosophy of Modern Physics\/}.

\bibitem[{Kraichnan({[1955]})}]{kraichnan1955}
Kraichnan, R.~H. {[1955]}:  `Special-Relativistic Derivation of Generally
  Covariant Gravitation Theory',
\newblock {\em Phys. Rev.\/}, {\textbf{ 98}\/}, pp. 1118--1122.

\bibitem[{Kraus({[1953]})}]{kraus1953}
Kraus, J.~D. {[1953]}: {\em Electromagnetics\/},
\newblock McGraw-Hill.

\bibitem[{Lam({[2011]})}]{lam2011}
Lam, V. {[2011]}:  `Gravitational and Nongravitational Energy: The Need for
  Background Structures',
\newblock {\em Philosophy of Science\/}, {\textbf{ 78}\/}(5), pp. 1012--1024.

\bibitem[{Landau and Lifshitz({[1972]})}]{landaulifshitzfields}
Landau, L.  and Lifshitz, E. {[1972]}: {\em Course of Theoretical Physics,
  Volume 2: The Classical Theory of Fields\/},
\newblock Addison-Wesley Publishing Company, 3rd edition.

\bibitem[{Lange({[2001]})}]{lange2001}
Lange, M. {[2001]}:  `The Most Famous Equation',
\newblock {\em The Journal of Philosophy\/}, {\textbf{ 98}\/}(5), pp. 219--238.

\bibitem[{Lange({[2002]})}]{lange}
Lange, M. {[2002]}: {\em An Introduction to the Philosophy of Physics:
  Locality, Energy, Fields, and Mass\/},
\newblock Blackwell.

\bibitem[{Lazarovici({[2017]})}]{Lazarovici2017}
Lazarovici, D. {[2017]}:  `Against Fields',
\newblock {\em European Journal for Philosophy of Science\/}.

\bibitem[{Lehmkuhl({[2008]})}]{lehmkuhl2008}
Lehmkuhl, D. {[2008]}:  `Is Spacetime a Gravitational Field?',
\newblock in D.~Dieks (\emph{ed.}), {\em The Ontology of Spacetime II\/},
  Elsevier, pp. 83--110.

\bibitem[{Lehmkuhl({[2011]})}]{lehmkuhl2011}
Lehmkuhl, D. {[2011]}:  `Mass-Energy-Momentum: Only there Because of
  Spacetime?',
\newblock {\em The British Journal for the Philosophy of Science\/}, {\textbf{
  62}\/}(3), pp. 453--488.

\bibitem[{Maartens and Bassett({[1998]})}]{maartens1998}
Maartens, R.  and Bassett, B.~A. {[1998]}:  `Gravito-electromagnetism',
\newblock {\em Classical and Quantum Gravity\/}, {\textbf{ 15}\/}(3), pp.
  705--717.

\bibitem[{Mashhoon({[2001]})}]{mashhoon2001b}
Mashhoon, B. {[2001]}:  `Gravitoelectromagnetism',
\newblock in J.~F. Pascual-S\'{a}nchez, L.~Flor\'{i}a, A.~S. Miguel  and
  F.~Vicente (\emph{eds}), {\em Reference Frames and Gravitomagnetism\/}, World
  Scientific, pp. 121--132.

\bibitem[{Mashhoon({[2007]})}]{mashhoon2007}
Mashhoon, B. {[2007]}:  `Gravitoelectromagnetism: A Brief Review',
\newblock in L.~Iorio (\emph{ed.}), {\em The Measurement of Gravitomagnetism: A
  Challenging Enterprise\/}, Nova Science Publishers, pp. 27--39.

\bibitem[{Mashhoon \emph{et~al.}({[2001]})Mashhoon, Gronwald, and
  Lichtenegger}]{mashhoon2001}
Mashhoon, B., Gronwald, F.  and Lichtenegger, H.~I. {[2001]}:
  `Gravitomagnetism and the Clock Effect',
\newblock in C.~L\"{a}mmerzahl, C.~Everitt  and F.~Hehl (\emph{eds}), {\em
  Gyros, Clocks, Interferometers...: Testing Relativistic Gravity in Space\/},
  Springer, pp. 83--108.

\bibitem[{Mashhoon \emph{et~al.}({[1999]})Mashhoon, McClune, and
  Quevedo}]{mashhoon1999}
Mashhoon, B., McClune, J.~C.  and Quevedo, H. {[1999]}:  `On the
  Gravitoelectromagnetic Stress-Energy Tensor',
\newblock {\em Classical and Quantum Gravity\/}, {\textbf{ 16}\/}(4), pp.
  1137--1148.

\bibitem[{Maudlin({[1988]})}]{maudlin1988}
Maudlin, T. {[1988]}:  `The Essence of Space-Time',
\newblock {\em PSA: Proceedings of the Biennial Meeting of the Philosophy of
  Science Association 1988\/}, {\textbf{ 2}\/}, pp. 82--91.

\bibitem[{Maxwell({[1864]})}]{maxwellgravity}
Maxwell, J.~C. {[1864]}:  `A Dynamical Theory of the Electromagnetic Field',
\newblock {\em Royal Society Transactions\/}, {\textbf{ CLV}\/},
\newblock reprinted in pp.\ 526--597 of \citep{maxwellvol}.

\bibitem[{Maxwell({[1890]})}]{maxwellvol}
Maxwell, J.~C. {[1890]}: {\em The Scientific Papers of James Clerk Maxwell,
  Volume 1\/},
\newblock W. D. Niven (\emph{ed.}), Cambridge University Press.

\bibitem[{Misner \emph{et~al.}({[1973]})Misner, Thorne, and Wheeler}]{MTW}
Misner, C.~W., Thorne, K.~S.  and Wheeler, J.~A. {[1973]}: {\em Gravitation\/},
\newblock W.~H. Freeman and Company.

\bibitem[{Nikishov({[2001]})}]{nikishov2001}
Nikishov, A.~I. {[2001]}:  `On Energy Momentum Tensors of Gravitational Field',
\newblock {\em Physics of Particles and Nuclei\/}, {\textbf{ 32}\/}, pp. 1--14.

\bibitem[{Noonan({[1984]})}]{noonan1984}
Noonan, T.~W. {[1984]}:  `The Gravitational Contribution to the Stress-Energy
  Tensor of a Medium in General Relativity',
\newblock {\em General Relativity and Gravitation\/}, {\textbf{ 16}\/}(11), pp.
  1103--1118.

\bibitem[{Ohanian({[unpublished]})}]{ohanian2010}
Ohanian, H.~C. {[unpublished]}:  `The Energy-Momentum Tensor in General
  Relativity and in Alternative Theories of Gravitation, and the Gravitational
  vs. Inertial Mass',
\newblock available at
  $\langle$\href{https://arxiv.org/abs/1010.5557}{arxiv.org/abs/1010.5557}$\rangle$.

\bibitem[{Ohanian and Ruffini({[2013]})}]{ohanianGR}
Ohanian, H.~C.  and Ruffini, R. {[2013]}: {\em Gravitation and Spacetime\/},
\newblock Cambridge University Press, 3rd edition.

\bibitem[{Okun({[1989]})}]{okun1989}
Okun, L.~B. {[1989]}:  `The Concept of Mass',
\newblock {\em Physics Today\/}, {\textbf{ 42}\/}(6), pp. 31--36.

\bibitem[{Padmanabhan({[2008]})}]{padmanabhan}
Padmanabhan, T. {[2008]}:  `From Gravitons to Gravity: Myths and Reality',
\newblock {\em International Journal of Modern Physics D\/}, {\textbf{
  17}\/}(03n04), pp. 367--398.

\bibitem[{Pearle({[1982]})}]{pearle1982}
Pearle, P. {[1982]}:  `Classical Electron Models',
\newblock in D.~Teplitz (\emph{ed.}), {\em Electromagnetism: Paths to
  Research\/}, Plenum Press, pp. 211--295.

\bibitem[{Peters({[1981]})}]{peters1981}
Peters, P.~C. {[1981]}:  `Where is the Energy Stored in a Gravitational
  Field?',
\newblock {\em American Journal of Physics\/}, {\textbf{ 49}\/}(6), pp.
  564--569.

\bibitem[{Pinto-Neto and Trajtenberg({[2000]})}]{pintoneto2000}
Pinto-Neto, N.  and Trajtenberg, P.~I. {[2000]}:  `On the Localization of the
  Gravitational Energy',
\newblock {\em Brazilian Journal of Physics\/}, {\textbf{ 30}\/}(1), pp.
  181--188.

\bibitem[{Pitts({[2010]})}]{pitts2010}
Pitts, J.~B. {[2010]}:  `Gauge-Invariant Localization of Infinitely Many
  Gravitational Energies from All Possible Auxiliary Structures',
\newblock {\em General Relativity and Gravitation\/}, {\textbf{ 42}\/}(3), pp.
  601--622.

\bibitem[{Pitts and Schieve({[2007]})}]{pittsschieve2007}
Pitts, J.~B.  and Schieve, W. {[2007]}:  `Universally Coupled Massive Gravity',
\newblock {\em Theoretical and Mathematical Physics\/}, {\textbf{ 151}\/}(2),
  pp. 700--717.

\bibitem[{Poincar\'{e}({[1900]})}]{poincare1900}
Poincar\'{e}, H. {[1900]}:  `La Th\'{e}orie de Lorentz et le Principe de
  R\'{e}action',
\newblock {\em Archives N\'{e}erlandaises des Sciences Exactes et
  Naturelles\/}, {\textbf{ 5}\/}, pp. 252--278,\
\newblock translation by S.~Lawrence available at
  $\langle$\url{http://www.physicsinsights.org/poincare-1900.pdf}$\rangle$.

\bibitem[{Poisson and Will({[2014]})}]{poissonwill}
Poisson, E.  and Will, C.~M. {[2014]}: {\em Gravity: Newtonian, Post-Newtonian,
  Relativistic\/},
\newblock Cambridge University Press.

\bibitem[{Purcell and Morin({[2013]})}]{purcell}
Purcell, E.~M.  and Morin, D.~J. {[2013]}: {\em Electricity and Magnetism\/},
\newblock Cambridge University Press, 3rd edition.

\bibitem[{Read({[forthcoming]})}]{readforth}
Read, J. {[forthcoming]}:  `Functional Gravitational Energy',
\newblock {\em The British Journal for the Philosophy of Science\/}.

\bibitem[{Rey({[unpublished]})}]{rey2013}
Rey, D. {[unpublished]}:  `Similarity Assessments, Spacetime, and the
  Gravitational Field: What does the Metric Tensor Represent in General
  Relativity?',
\newblock available at
  $\langle$\href{http://philsci-archive.pitt.edu/9615/}{philsci-archive.pitt.edu/9615/}$\rangle$.

\bibitem[{Rindler({[2006]})}]{rindler2006}
Rindler, W. {[2006]}: {\em Relativity: Special, General, and Cosmological\/},
\newblock Oxford University Press, 2nd edition.

\bibitem[{Rindler \emph{et~al.}({[1990]})Rindler, Vandyck, Murugesan, Ruschin,
  Sauter, and Okun}]{rindler1990}
Rindler, W., Vandyck, M.~A., Murugesan, P., Ruschin, S., Sauter, C.  and Okun,
  L.~B. {[1990]}:  `Putting to Rest Mass Misconceptions',
\newblock {\em Physics Today\/}, {\textbf{ 43}\/}(5), pp. 13--15, 115--117.

\bibitem[{Roller and Roller({[1954]})}]{roller1954}
Roller, D.  and Roller, D. H.~D. {[1954]}: {\em The Development of the Concept
  of Electric Charge: Electricity from the Greeks to Coulomb\/},
\newblock Harvard University Press.

\bibitem[{Sebens({[2018]})}]{forcesonfields}
Sebens, C.~T. {[2018]}:  `Forces on Fields',
\newblock {\em Studies in History and Philosophy of Modern Physics\/},
  {\textbf{ 63}\/}, pp. 1--11.

\bibitem[{Sebens({[unpublished]})}]{howelectronsspin}
Sebens, C.~T. {[unpublished]}:  `How Electrons Spin',
\newblock available at
  $\langle$\href{https://arxiv.org/abs/1806.01121}{arxiv.org/abs/1806.01121}$\rangle$.

\bibitem[{Straumann({[2004]})}]{straumann2004}
Straumann, N. {[2004]}: {\em General Relativity With Applications to
  Astrophysics\/},
\newblock Springer.

\bibitem[{Straumann({[unpublished]})}]{straumann2000}
Straumann, N. {[unpublished]}:  `Reflections on Gravity',
\newblock available at
  $\langle$\href{https://arxiv.org/abs/astro-ph/0006423}{arxiv.org/abs/astro-ph/0006423}$\rangle$.

\bibitem[{Synge({[1972]})}]{synge1972}
Synge, J.~L. {[1972]}:  `Newtonian Gravitational Field Theory',
\newblock {\em Il Nuovo Cimento B\/}, {\textbf{ 8}\/}(2), pp. 373--390.

\bibitem[{Szabados({[2009]})}]{szabados2009}
Szabados, L.~B. {[2009]}:  `Quasi-Local Energy-Momentum and Angular Momentum in
  General Relativity',
\newblock {\em Living Reviews in Relativity\/}, {\textbf{ 12}\/}, pp. 4.

\bibitem[{Tartaglia and Ruggiero({[2003]})}]{tartaglia2003}
Tartaglia, A.  and Ruggiero, M.~L. {[2003]}:  `Gravito-Electromagnetism Versus
  Electromagnetism',
\newblock {\em European Journal of Physics\/}, {\textbf{ 25}\/}(2), pp.
  203--210.

\bibitem[{Taylor and Wheeler({[1992]})}]{taylorwheeler}
Taylor, E.~F.  and Wheeler, J.~A. {[1992]}: {\em Spacetime Physics:
  Introduction to Special Relativity\/},
\newblock W.H. Freeman and Company, 2nd edition.

\bibitem[{Thirring({[1961]})}]{thirring1961}
Thirring, W.~E. {[1961]}:  `An Alternative Approach to the Theory of
  Gravitation',
\newblock {\em Annals of Physics\/}, {\textbf{ 16}\/}(1), pp. 96--117.

\bibitem[{Thorne and Blandford({[2017]})}]{thorneblandford}
Thorne, K.~S.  and Blandford, R.~D. {[2017]}: {\em Modern Classical Physics\/},
\newblock Princeton University Press.

\bibitem[{Visser({[1989]})}]{visser1989}
Visser, M. {[1989]}:  `A Classical Model for the Electron',
\newblock {\em Physics Letters A\/}, {\textbf{ 139}\/}(3), pp. 99--102.

\bibitem[{Wald({[1984]})}]{waldGR}
Wald, R.~M. {[1984]}: {\em General Relativity\/},
\newblock University of Chicago Press.

\bibitem[{Wallace({[2012]})}]{wallace2012}
Wallace, D. {[2012]}: {\em The Emergent Multiverse: Quantum Theory According to
  the Everett Interpretation\/},
\newblock Oxford University Press.

\bibitem[{Weinberg({[1972]})}]{weinbergGR}
Weinberg, S. {[1972]}: {\em Gravitation and Cosmology: Principles and
  Applications of the General Theory of Relativity\/},
\newblock Wiley.

\bibitem[{Whittaker({[1951]})}]{whittaker1}
Whittaker, E.~T. {[1951]}: {\em A History of the Theories of Aether \&
  Electricity\/},
\newblock vol.~1
\newblock Thomas Nelson \& Sons.

\end{thebibliography}
\end{document}